\documentstyle[aps,epsfig,floats]{revtex}

\tighten
\begin{document}

\draft

\title{
\begin{flushright}
\begin{minipage}{3 cm}
\small
hep-ph/9704335\\
NIKHEF 97-018\\ 
VUTH 97-7
\end{minipage}
\end{flushright}
Modelling quark distribution and fragmentation functions}

\author{R. Jakob$^1$, P.J. Mulders$^{1,2}$ and J. Rodrigues$^{1,3}$}
\address{\mbox{}\\
$^1$National Institute for Nuclear Physics and High--Energy
Physics (NIKHEF)\\
P.O. Box 41882, NL-1009 DB Amsterdam, the Netherlands\\
\mbox{}\\
$^2$Department of Physics and Astronomy, Free University \\
De Boelelaan 1081, NL-1081 HV Amsterdam, the Netherlands \\
\mbox{}\\
$^3$Instituto Superior T\'{e}cnico \\
Av. Rovisco Pais, 1100 Lisboa Codex, Portugal\\[6mm]}

\date{April 15, 1997}

\maketitle

\begin{abstract}
The representation of quark distribution and fragmentation functions in
terms of non-local operators is combined with a simple spectator model.
This allows us to estimate these functions for the nucleon and the pion
ensuring correct crossing and support properties. We give estimates for
the unpolarized functions as well as for the polarized ones and for subleading 
(higher twist) functions. Furthermore we can study several relations that
are consequences of Lorentz invariance and of C, P, and T invariance of the
strong interactions. 
\end{abstract}



\newcommand{\bm}[1]{\mbox{\boldmath $#1$}}
\newcommand{\tv}[1]{{\bm #1}_{T}}
\newcommand{\ptinst}{\,\tv{p}\!\cdot\!\tv{S}\,}
\newcommand{\ktinsht}{\,\tv{k}\!\cdot\!{\bm S}_{hT}\,}
\newcommand{\nnn}{\nonumber\\}
\newcommand{\com}[1]{($\clubsuit$~{\em #1}~!?!)}
\newcommand{\tstrut}{\rule[-1.6mm]{0mm}{5.6mm}}


\newcommand{\be}{\begin{equation}}
\newcommand{\bea}{\begin{eqnarray}}
\newcommand{\ee}{\end{equation}}
\newcommand{\eea}{\end{eqnarray}}
\newcommand{\ba}{\begin{array}}
\newcommand{\ea}{\end{array}}
\newcommand{\nin}{\noindent}
\newcommand{\nn}{\nonumber}

\newcommand{\larrow}{\leftarrow}
\newcommand{\rarrow}{\rightarrow}
\newcommand{\darrow}{\downarrow}
\newcommand{\lrarrow}{\leftrightarrow}

\newcommand{\tr}{{\rm Tr}}


\newcommand{\ar}{a_{\scriptscriptstyle R}}
\newcommand{\epsR}{\epsilon^{\scriptscriptstyle R}}
\newcommand{\gA}{g_{\scriptscriptstyle A}}
\newcommand{\gr}{g_{\scriptscriptstyle R}}
\newcommand{\lamR}{\lambda_{\scriptscriptstyle R}}
\newcommand{\xb}{x_{\scriptscriptstyle B}}


\newcommand{\half}{{1 \over 2}}


\newcommand{\dk}{\int {d^4k \over {(2 \pi)}^4} \ }
\newcommand{\dl}{\int {d^dl \over {(2 \pi)}^4} \ }


\newcommand{\smn}{\sigma^{\mu \nu}}


\newcommand{\lc}{\epsilon_{\mu \nu \rho \sigma}}
\newcommand{\lcup}{\epsilon^{\mu \nu \rho \sigma}}
\newcommand{\lcperp}{\epsilon^{\mu \nu}_\perp}


\newcommand{\ga}{\gamma_{5}}
\newcommand{\gp}{\gamma^{+}}
\newcommand{\gm}{\gamma^{-}}


\newcommand{\gma}{g^{\mu \alpha}}
\newcommand{\gmb}{g^{\mu \beta}}
\newcommand{\gmg}{g^{\mu \gamma}}
\newcommand{\gmn}{g^{\mu \nu}}
\newcommand{\gmr}{g^{\mu \rho}}
\newcommand{\gms}{g^{\mu \sigma}}

\newcommand{\gmnperp}{g^{\mu \nu}_\perp}


\newcommand{\dalpha}{\partial_\alpha}
\newcommand{\dbeta}{\partial_\beta}
\newcommand{\dmu}{\partial_\mu}
\newcommand{\dnu}{\partial_\nu}
\newcommand{\drho}{\partial_\rho}
\newcommand{\Dmu}{D_\mu}

\newcommand{\dcalpha}{\partial^\alpha}
\newcommand{\dcbeta}{\partial^\beta}
\newcommand{\dcmu}{\partial^\mu}
\newcommand{\dcnu}{\partial^\nu}
\newcommand{\dcrho}{\partial^\rho}


\def\sla#1{\setbox0=\hbox{$#1$}
   \dimen0=\wd0 \setbox1=\hbox{/} \dimen1=\wd1
   \ifdim\dimen0>\dimen1 \rlap{\hbox to \dimen0{\hfil/\hfil}} #1
   \else  \rlap{\hbox to \dimen1{\hfil$#1$\hfil}} / \fi}


\newcommand{\alp}{\alpha}
\newcommand{\bet}{\beta}
\newcommand{\gamm}{\gamma}
\newcommand{\Gam}{\Gamma}
\newcommand{\Gamm}{\Gamma}
\newcommand{\del}{\delta}
\newcommand{\Del}{\Delta}
\newcommand{\eps}{\epsilon}
\newcommand{\lam}{\lambda}
\newcommand{\Lam}{\Lambda}
\newcommand{\sig}{\sigma}
\newcommand{\Sig}{\Sigma}


\newcommand{\kt}{{\bm k}_T^2}
\newcommand{\kts}{{\bm k}_T^2}
\newcommand{\ktns}{{\bm k}_T}
\newcommand{\kp}{{{\bm k}'_T}^2}
\newcommand{\kpns}{{\bm k}'_T}
\newcommand{\qt}{{\bm q}_T^2}
\newcommand{\qtns}{{\bm q}_T}
\newcommand{\pt}{{\bm p}_T^2}
\newcommand{\pts}{{\bm p}_T^2}
\newcommand{\ptns}{{\bm p}_T}
\newcommand{\st}{{\bm S}_T^2}
\newcommand{\stns}{{\bm S}_T}
\newcommand{\sht}{{\bm S}_{hT}^2}
\newcommand{\shtns}{{\bm S}_{hT}}
\newcommand{\xpts}{(x,{\bm p}_T^2)}
\newcommand{\xptns}{(x,{\bm p}_T)}
\newcommand{\zkp}{(z,{{\bm k}'_T}^2)}
\newcommand{\zkpns}{(z,{\bm k}'_T)}
\newcommand{\zkt}{(z,{{\bm k}_T}^2)}
\newcommand{\zktns}{(z,{\bm k}_T)}


\newcommand{\cfa}{\Phi^{[\Gamma]} (x,{\bm p}_T)}
\newcommand{\cfgp}{\Phi^{[\gamma^+]} (x,{\bm p}_T)}
\newcommand{\cfgpga}{\Phi^{[\gamma^+ \gamma^5]} (x,{\bm p}_T)}
\newcommand{\cfsa}{\Phi^{[i\sigma^{i+} \gamma^5]} (x,{\bm p}_T)}


\newcommand{\fo}{f_1}
\newcommand{\fox}{f_1(x)}
\newcommand{\foxk}{f_1(x,{\bm k}^2_T)}
\newcommand{\foxp}{f_1(x,{\bm p}^2_T)}

\newcommand{\go}{g_1}
\newcommand{\gol}{g_{1L}}
\newcommand{\got}{g_{1T}}
\newcommand{\goto}{g_{1T}^{(1)}}
\newcommand{\gos}{g_{1s}}
\newcommand{\gox}{g_{1}(x)}
\newcommand{\gotox}{g_{1T}^{(1)}(x)}
\newcommand{\golxk}{g_{1L}(x,{\bm k}^2_T)}
\newcommand{\golxp}{g_{1L}(x,{\bm p}^2_T)}
\newcommand{\gotxk}{g_{1T}(x,{\bm k}^2_T)}
\newcommand{\gotxp}{g_{1T}(x,{\bm p}^2_T)}
\newcommand{\gosxk}{g_{1s}(x,{\bm k}^2_T)}
\newcommand{\gosxp}{g_{1s}(x,{\bm p}^2_T)}
\newcommand{\gsperp}{g_s^\perp}
\newcommand{\gtwo}{g_2}
\newcommand{\gtwox}{g_2(x)}

\newcommand{\ho}{h_{1}}
\newcommand{\hox}{h_{1}(x)}
\newcommand{\hot}{h_{1T}}
\newcommand{\hotxk}{h_{1T}(x,{\bm k}^2_T)}
\newcommand{\hotxp}{h_{1T}(x,{\bm p}^2_T)}
\newcommand{\holperp}{h_{1L}^\perp}
\newcommand{\hotperp}{h_{1T}^\perp}
\newcommand{\hosperp}{h_{1s}^\perp}
\newcommand{\holperpo}{h_{1L}^{\perp (1)}}
\newcommand{\holperpox}{h_{1L}^{\perp (1)}(x)}
\newcommand{\hotperpt}{h_{1T}^{\perp (2)}}
\newcommand{\hotperptx}{h_{1T}^{\perp (2)}(x)}
\newcommand{\holperpxk}{h_{1L}^\perp (x,{\bm k}^2_T)}
\newcommand{\holperpxp}{h_{1L}^\perp (x,{\bm p}^2_T)}
\newcommand{\hotperpxk}{h_{1T}^\perp (x,{\bm k}^2_T)}
\newcommand{\hotperpxp}{h_{1T}^\perp (x,{\bm p}^2_T)}
\newcommand{\hosperpxk}{h_{1s}^\perp (x,{\bm k}^2_T)}
\newcommand{\hosperpxp}{h_{1s}^\perp (x,{\bm p}^2_T)}
\newcommand{\htwo}{h_2}
\newcommand{\htwox}{h_2(x)}

\newcommand{\ex}{e(x)}
\newcommand{\exk}{e(x,{\bm k}^2_T)}
\newcommand{\expp}{e(x,{\bm p}^2_T)}

\newcommand{\fperp}{f^\perp}
\newcommand{\fperpxk}{f^\perp (x,{\bm k}^2_T)}
\newcommand{\fperpxp}{f^\perp (x,{\bm p}^2_T)}

\newcommand{\gprimet}{g'_T}
\newcommand{\gprimetxk}{g'_T (x,{\bm k}^2_T)}
\newcommand{\gprimetxp}{g'_T (x,{\bm p}^2_T)}

\newcommand{\glperp}{g_L^\perp}
\newcommand{\gtperp}{g_T^\perp}
\newcommand{\glperpo}{g_L^{\perp (1)}}
\newcommand{\glperpox}{g_L^{\perp (1)}(x)}
\newcommand{\gtperpt}{g_T^{\perp (2)}}
\newcommand{\gtperptx}{g_T^{\perp (2)}(x)}
\newcommand{\glperpxk}{g_L^\perp (x,{\bm k}^2_T)}
\newcommand{\glperpxp}{g_L^\perp (x,{\bm p}^2_T)}
\newcommand{\gtperpxk}{g_T^\perp (x,{\bm k}^2_T)}
\newcommand{\gtperpxp}{g_T^\perp (x,{\bm p}^2_T)}
\newcommand{\gsperpxk}{g_s^\perp (x,{\bm k}^2_T)}
\newcommand{\gsperpxp}{g_s^\perp (x,{\bm p}^2_T)}

\newcommand{\gt}{g_T}
\newcommand{\gtx}{g_T(x)}
\newcommand{\hl}{h_L}
\newcommand{\hs}{h_s}
\newcommand{\hlx}{h_L(x)}

\newcommand{\htperp}{h_T^\perp}
\newcommand{\htperpxk}{h_T^\perp (x,{\bm k}^2_T)}
\newcommand{\htperpxp}{h_T^\perp (x,{\bm p}^2_T)}
\newcommand{\htt}{h_T}
\newcommand{\htxk}{h_T (x,{\bm k}^2_T)}
\newcommand{\htxp}{h_T (x,{\bm p}^2_T)}
\newcommand{\hlxk}{h_L (x,{\bm k}^2_T)}
\newcommand{\hlxp}{h_L (x,{\bm p}^2_T)}
\newcommand{\hsxk}{h_s (x,{\bm k}^2_T)}
\newcommand{\hsxp}{h_s (x,{\bm p}^2_T)}
\newcommand{\hto}{h_T^{(1)}}
\newcommand{\htox}{h_T^{(1)}(x)}
\newcommand{\htperpo}{h_T^{\perp (1)}}
\newcommand{\htperpox}{h_T^{\perp (1)}(x)}


\newcommand{\Do}{D_1}
\newcommand{\Doz}{D_{1}(z)}
\newcommand{\Dozk}{D_{1}(z,{\bm k}^2_T)}
\newcommand{\Dozp}{D_{1}(z,{\bm p}^2_T)}

\newcommand{\Dotperp}{D_{1T}^\perp}
\newcommand{\Dotperpzk}{D_{1T}^\perp (z,{\bm k}^2_T)}
\newcommand{\Dotperpzp}{D_{1T}^\perp (z,{\bm p}^2_T)}
\newcommand{\Dotperpo}{D_{1T}^{\perp(1)}}
\newcommand{\Dotperpoz}{D_{1T}^{\perp(1)}(z)}

\newcommand{\Go}{G_{1}}
\newcommand{\Goz}{G_{1}(z)}

\newcommand{\Gol}{G_{1L}}
\newcommand{\Golzk}{G_{1L}(z,{\bm k}^2_T)}
\newcommand{\Golzp}{G_{1L}(z,{\bm p}^2_T)}

\newcommand{\Got}{G_{1T}}
\newcommand{\Gotzk}{G_{1T}(z,{\bm k}^2_T)}
\newcommand{\Gotzp}{G_{1T}(z,{\bm p}^2_T)}
\newcommand{\Goto}{G_{1T}^{(1)}}
\newcommand{\Gotoz}{G_{1T}^{(1)}(z)}

\newcommand{\Gos}{G_{1s}}
\newcommand{\Goszk}{G_{1s}(z,{\bm k}^2_T)}
\newcommand{\Goszp}{G_{1s}(z,{\bm p}^2_T)}

\newcommand{\Gtwo}{G_2}
\newcommand{\Gtwoz}{G_2(z)}

\newcommand{\Ho}{H_{1}}
\newcommand{\Hoz}{H_{1}(z)}

\newcommand{\Hot}{H_{1T}}
\newcommand{\Hotzk}{H_{1T}(z,{\bm k}^2_T)}
\newcommand{\Hotzp}{H_{1T}(z,{\bm p}^2_T)}

\newcommand{\Hoperp}{H_1^\perp}
\newcommand{\Hoperpzk}{H_1^\perp (z,{\bm k}^2_T)}
\newcommand{\Hoperpzp}{H_1^\perp (z,{\bm p}^2_T)}
\newcommand{\Hoperpo}{H_1^{\perp(1)}}
\newcommand{\Hoperpoz}{H_1^{\perp(1)}(z)}

\newcommand{\Holperp}{H_{1L}^\perp}
\newcommand{\Holperpzk}{H_{1L}^\perp (z,{\bm k}^2_T)}
\newcommand{\Holperpzp}{H_{1L}^\perp (z,{\bm p}^2_T)}
\newcommand{\Holperpo}{H_{1L}^{\perp(1)}}
\newcommand{\Holperpoz}{H_{1L}^{\perp(1)}(z)}

\newcommand{\Hotperp}{H_{1T}^\perp}
\newcommand{\Hotperpzk}{H_{1T}^\perp (z,{\bm k}^2_T)}
\newcommand{\Hotperpzp}{H_{1T}^\perp (z,{\bm p}^2_T)}
\newcommand{\Hotperpt}{H_{1T}^{\perp(2)}}
\newcommand{\Hotperptz}{H_{1T}^{\perp(2)}(z)}

\newcommand{\Hosperp}{H_{1s}^\perp}
\newcommand{\Hosperpzk}{H_{1s}^\perp (z,{\bm k}^2_T)}
\newcommand{\Hosperpzp}{H_{1s}^\perp (z,{\bm p}^2_T)}

\newcommand{\Htwo}{H_2}
\newcommand{\Htwoz}{H_2(z)}

\newcommand{\Ez}{E(z)}
\newcommand{\Ezk}{E(z,{\bm k}^2_T)}
\newcommand{\Ezp}{E(z,{\bm p}^2_T)}

\newcommand{\El}{E_L}
\newcommand{\Elz}{E_L(z)}
\newcommand{\Elzk}{E_L (z,{\bm k}^2_T)}
\newcommand{\Elzp}{E_L (z,{\bm p}^2_T)}

\newcommand{\Et}{E_T}
\newcommand{\Etz}{E_T(z)}
\newcommand{\Etzk}{E_T (z,{\bm k}^2_T)}
\newcommand{\Etzp}{E_T (z,{\bm p}^2_T)}
\newcommand{\Etoz}{E_T^{(1)}(z)}

\newcommand{\Dperp}{D^\perp}
\newcommand{\Dperpzk}{D^\perp (z,{\bm k}^2_T)}
\newcommand{\Dperpzp}{D^\perp (z,{\bm p}^2_T)}

\newcommand{\Dlperp}{D_L^\perp}
\newcommand{\Dlperpzk}{D_L^\perp (z,{\bm k}^2_T)}
\newcommand{\Dlperpzp}{D_L^\perp (z,{\bm p}^2_T)}
\newcommand{\Dlperpo}{D_L^{\perp(1)}}
\newcommand{\Dlperpoz}{D_L^{\perp(1)}(z)}

\newcommand{\Dt}{D_T}
\newcommand{\Dtz}{D_T(z)}
\newcommand{\Dtzk}{D_T (z,{\bm k}^2_T)}
\newcommand{\Dtzp}{D_T (z,{\bm p}^2_T)}

\newcommand{\Gprimet}{G'_T}
\newcommand{\Gprimetzk}{G'_T (z,{\bm k}^2_T)}
\newcommand{\Gprimetzp}{G'_T (z,{\bm p}^2_T)}

\newcommand{\Glperp}{G_L^\perp}
\newcommand{\Glperpzk}{G_L^\perp (z,{\bm k}^2_T)}
\newcommand{\Glperpzp}{G_L^\perp (z,{\bm p}^2_T)}
\newcommand{\Glperpo}{G_L^{\perp(1)}}
\newcommand{\Glperpoz}{G_L^{\perp(1)}(z)}

\newcommand{\Gtperp}{G_T^\perp}
\newcommand{\Gtperpzk}{G_T^\perp (z,{\bm k}^2_T)}
\newcommand{\Gtperpzp}{G_T^\perp (z,{\bm p}^2_T)}
\newcommand{\Gtperpt}{G_T^{\perp(2)}}
\newcommand{\Gtperptz}{G_T^{\perp(2)}(z)}

\newcommand{\Gsperpzk}{G_s^\perp (z,{\bm k}^2_T)}
\newcommand{\Gsperpzp}{G_s^\perp (z,{\bm p}^2_T)}

\newcommand{\Gt}{G_T}
\newcommand{\Gtz}{G_T(z)}

\newcommand{\Hz}{H(z)}
\newcommand{\Hzk}{H(z,{\bm k}^2_T)}

\newcommand{\Hl}{H_L}
\newcommand{\Hlz}{H_L(z)}
\newcommand{\Hlzk}{H_L (z,{\bm k}^2_T)}
\newcommand{\Hlzp}{H_L (z,{\bm p}^2_T)}

\newcommand{\Ht}{H_T}
\newcommand{\Htz}{H_T(z)}
\newcommand{\Htzk}{H_T (z,{\bm k}^2_T)}
\newcommand{\Htzp}{H_T (z,{\bm p}^2_T)}
\newcommand{\Hto}{H_T^{(1)}}
\newcommand{\Htoz}{H_T^{(1)}(z)}

\newcommand{\Hs}{H_s}
\newcommand{\Hszk}{H_s (z,{\bm k}^2_T)}
\newcommand{\Hszp}{H_s (z,{\bm p}^2_T)}

\newcommand{\Htperp}{H_T^\perp}
\newcommand{\Htperpzk}{H_T^\perp (z,{\bm k}^2_T)}
\newcommand{\Htperpzp}{H_T^\perp (z,{\bm p}^2_T)}
\newcommand{\Htperpo}{H_T^{\perp(1)}}
\newcommand{\Htperpoz}{H_T^{\perp(1)}(z)}
\newcommand{\Htperpt}{H_T^{\perp(2)}}
\newcommand{\Htperptz}{H_T^{\perp(2)}(z)}


\newcommand{\bpsi}{\overline{\psi}}
\newcommand{\bpsix}{\bar{\psi}(x)}
\newcommand{\dsdt}{\int \ [d \sig d \tau]}
\newcommand{\dshdth}{\int \ [d \sig_h d \tau_h]}
\newcommand{\dshdthb}{\int \ \{d \sig_h d \tau_h\}}
\newcommand{\hadront}{2MW^{\mu \nu }}
\newcommand{\mh}{M_h}


\section{Introduction}

\begin{figure}[ht]
\begin{center}
\epsfig{file=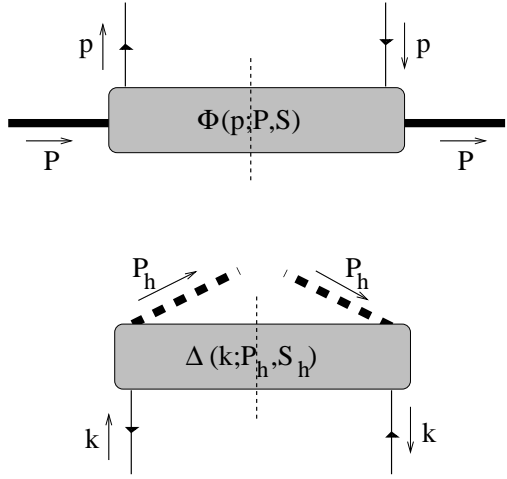,width=0.4\textwidth,angle=0}
\caption{\label{correlators} Diagrammatic representation of the 
correlation functions $\Phi$ and $\Delta$.}
\end{center}
\end{figure}

Quark distribution functions and quark fragmentation functions appear in
the field-theoretical description of hard processes as the parts that
connect the quark and gluon lines to hadrons in the initial or final state.
These parts are defined as connected matrix elements of non-local
operators built from quark and gluon fields. The simplest, but most important
ones, are the quark-quark correlation
functions\cite{Soper-77,Collins-Soper-82,Jaffe-83}.
For each quark flavor one can write
\bea
\Phi_{ij} (p, P, S) & = & \int {d^4 \xi \over (2 \pi)^4} \ e^{-i p \cdot \xi} \
\langle P, S | \bpsi_j(\xi) \psi_i(0) | P, S \rangle,
\label{qcf1}\\
\Del_{kl} (k, P_h, S_h) & = & \sum_X
\int {d^4 \xi \over (2 \pi)^4} \ e^{i k \cdot \xi} \
\langle 0 | \psi_k(\xi) | P_h, S_h; X \rangle \
\langle P_h, S_h; X | \bpsi_l(0) | 0 \rangle,
\label{qcf2}
\eea
diagrammatically represented in Fig.~\ref{correlators}. 
The hadron states are characterized by the momentum and spin vectors 
(limiting ourselves to spinless or spin 1/2 hadrons), $\vert P, S\rangle$ and
$\vert P_h, S_h\rangle$, for incoming and outgoing hadrons, respectively.
The quark momenta are denoted $p$ and $k$, respectively. In the correlation
function $\Delta$, the sum runs over all possible intermediate states
that contain the hadron $h$ characterized by $P_h$ and $S_h$, while for $\Phi$
the sum is omitted assuming completeness. Furthermore a summation (average)
over colors is understood in $\Phi$ ($\Delta$).

In a particular hard process only certain Dirac projections of the
correlation functions appear and the non-locality is restricted by the
integration over quark momenta. For inclusive lepton-hadron scattering,
where the hard scale $Q$ is set by the spacelike momentum transfer $-q^2$
= $Q^2$, the correlation function $\Phi$ appears in the leading order
result in an expansion in $1/Q$ of the cross section of hard processes. 
To be precise, the structure functions
appearing in the cross sections can be expressed in terms of quark distribution
functions, e.g., at leading order in $\alp_s$, $2F_1(\xb)$ = $F_2(\xb)/\xb$ =
$\sum_a e_a^2\,f_1^a(\xb)$, where $\xb = Q^2/2P\cdot q$.

The unpolarized quark distribution function 
(omitting flavor index $a$) is obtained from $\Phi$ as
\bea
f_1(x) & = & \left. \frac{1}{2}\int dp^-\,d^2\bm p_T\ \tr (\Phi\gamma^+) 
\right|_{p^+ = x P^+} \nn \\
& = & \left. \int \frac{d\xi^-}{4\pi}\ e^{i\,p^+\xi^-}
\langle P,S\vert \overline \psi (0)\gamma^+\psi(\xi)\vert P,S\rangle
\right|_{\xi^+ = \xi_T = 0},
\label{correlationf}
\eea
and depends only on the light-cone momentum fraction $x$ = $p^+/P^+$.
The lightlike components $a^\pm \equiv a\cdot n_\mp$ are defined with the
help of lightlike vectors $n_\pm$ satisfying $n_+^2$ = $n_-^2$ = 0 and
$n_+\cdot n_-$ = 1.
In inclusive lepton-hadron scattering they are related to the hadron 
momentum $P$ and the hard momentum $q$ as
\bea
P & = &
{M^2 x_b \over Q \sqrt{2}} \ n_- + {Q \over x_b \sqrt{2}} \ n_+,
\label{parp} \\
q & = & {Q \over \sqrt{2}} \ n_- - {Q \over \sqrt{2}} \ n_+,
\eea
\nin where $M$ is the mass of the hadron. 
The non-locality is restricted to a lightlike separation. At this point it
should be noted that there are other contributions in the leading
cross section resulting from soft parts with $A^+$ gluon legs. These can
be absorbed into the correlation function providing the link operator
that renders the definition in Eqs.~(\ref{qcf1}) and (\ref{qcf2}) color
gauge invariant. Choosing the $A^+ = 0$ gauge in the study of $\Phi$ the
link operator reduces to unity.

The simplest example of a hard process in which the correlation function
$\Delta$ appears is 1-particle inclusive $e^+e^-$ annihilation. In that
case the scale is set by the momentum squared of the annihilating
leptons, $q^2$ = $Q^2$ and the production cross section in leading order
becomes proportional to fragmentation functions
$D_1^{a\rightarrow h}(z_h)$, where $z_h$ = $2P_h\cdot q/q^2$.
The quark fragmentation function (omitting flavor index $a$) is obtained
from $\Delta$ as
\bea
D_1(z) & = & \left. \frac{1}{4z}\int dk^+\,d^2\bm k_T\ \tr (\Delta\gamma^-) 
\right|_{k^- = P_h^-/z} \nn \\
& = & \left. \int \frac{d\xi^+}{8\pi\,z}\ e^{i\,k^-\xi^+}\ \tr\left(
\gamma^- \langle 0\vert \psi (\xi)\vert P_h,S_h;X\rangle
\langle P_h,S_h;X\vert \overline \psi(0)\vert 0\rangle \right)
\right|_{\xi^- = \xi_T = 0},
\label{correlationD}
\eea
and depends only on the light-cone momentum fraction $z$ = $P_h^-/k^-$.
Taking as an explicit example 1-particle inclusive electron-positron 
annihilation, the lightlike vectors are defined from the hadron momentum 
$P_h$ and the hard momentum $q$:
\bea
P_h & = &{z_h Q \over \sqrt{2}} \ n_- + {\mh^2 \over z_h Q \sqrt{2}} \ n_+,
\label{parph} \\
q & = & {Q \over \sqrt{2}} \ n_- + {Q \over \sqrt{2}} \ n_+,
\eea
\nin where $\mh$ is the mass of the produced hadron.
The non-locality in the expression for $D_1$ is again restricted to a
lightlike separation. The link operator needed for color gauge invariance
becomes invisible by using the gauge $A^-$ = 0 in the study of $\Delta$.

As soon as in a hard process two hadrons (or for that matter one jet and
one hadron) play a role, the transverse directions become important.
Examples are inclusive Drell-Yan scattering, 1-particle inclusive
lepton-hadron scattering or 2-particle inclusive $e^+e^-$ annihilation.
For instance, in inclusive lepton-hadron scattering the leading order
cross section is given by the handbag diagram containing only one soft part
$\Phi$, shown in Fig.~\ref{correlators}, but in 1-particle inclusive 
lepton-hadron scattering the leading order cross section involves also the
fragmentation of the quark, the part $\Delta$ in Fig.~\ref{correlators}.
In order to describe the current fragmentation
one can still define lightlike vectors $n_\pm$ using the hadronic
momenta as in Eqs.~(\ref{parp}) and (\ref{parph}), but the third momentum,
in casu the hard momentum $q$, contains a transverse piece. Up to 
$M^2/Q^2$ corrections (irrelevant for our purposes), $P$ is proportional to
$n_+$, $P_h$ is proportional to $n_-$ and 
\be
q = {Q \over \sqrt{2}} \ n_- - {Q \over \sqrt{2}} \ n_+ + q_T.
\label{parq}
\ee
By selecting observables depending on the transverse momentum scale
$-q_T^2$ = $Q_T^2 \ll Q^2$ one needs to consider distribution
functions and fragmentation functions before integrating over $\bm p_T$
or $\bm k_T$, respectively. Examples are the dependence on the transverse
momentum of the produced lepton pair in Drell-Yan scattering, the
dependence on the transverse momentum of a produced hadron belonging to
the current jet in lepton-hadron scattering or the transverse momentum
distribution of hadrons with respect to the jet-axis (or with respect to a
fast hadron in the opposite jet) in the case of back-to-back jets in
$e^+e^-$ annihilation.

In this paper, we review the structure of light-cone correlation functions
including the effects of transverse separation of the quark fields,
and we estimate them using a simple model. This will be done for all
possible Dirac projections that contribute in leading or subleading order.
Thus we obtain estimates not only for the usual unpolarized distribution and
fragmentation functions, but also for the polarized ones and for subleading
(higher twist) functions.
As we will see, in a number of cases there are relations between 
leading and subleading $\bm p_T$-integrated functions.

Although hard cross sections can be expressed in terms of distribution
and fragmentation functions, these objects cannot be calculated from QCD
because they involve the hadronic bound states (at least not for light
quarks). Even the simplest case,
the moments that are related to matrix elements of local operators require
non-perturbative methods like, for instance, lattice calculations.
The full ($x$-dependent) quark distribution, however, requires the
knowledge of all moments.
We follow here a different route. We want to
investigate the structure of light-cone correlation functions and illustrate
the consequences of various constraints in a simple model. Particularly
suitable is a model in which the spectrum of intermediate states, which
can be inserted in Eq.~(\ref{qcf1}) 
or is explicitly present in Eq.~(\ref{qcf2}),
is replaced by one state, referred to as a {\em diquark}, 
if the hadron is a baryon.
At that point one still has lots of freedom to parametrize the 
hadron-quark-diquark vertex. 
We make the ansatz that in the zero-binding limit
implies the simple symmetric $SU(6)$ spin-flavor structure. As parameters
to describe the vertex one then has only the quark mass, a diquark mass and a
size parameter left. 

Although there are a few parameters, the approach has the advantage of 
being covariant and producing the right support, in contrast to the use
of other models \cite{song94}, such as bag 
models \cite{jaffe75,benesh88,schreiber91}, quark 
models \cite{ropele95,roberts96,barone97} or soliton 
models \cite{weigel96a,weigel96b,diakonov97}, which require projection 
techniques \cite{benesh87}. Of course, these models have the advantage that
they also reproduce other observables. Similar spectator models have also
been used for the calculation of distributions \cite{meyer90,joao95} 
and fragmentation
functions \cite{hoodbhoy95}.

Of course, matrix elements as discussed above have a scale dependence,
which in expressions for hard cross sections shows up as a logarithmic
dependence on the hard scale. A well known problem is that the modelling
does not provide a scale dependence. Models with a simple valence quark input, 
as we take as our starting point, will be naturally `low scale models'.
In principle, this could be evolved to higher scales to allow comparison
with data. However, we lack evolution equations at low scales. On the other
hand, we also consider higher twist distributions, for which
evolution is much more involved than the twist two case \cite{koike97}. 
These are the reasons for not including evolution in this paper. 
In other words, radiative corrections
and their absorption in the scale dependence of the quark distributions
are not taken into account.

The setup of the paper is the following. In section II we discuss the
structure of light-cone correlation functions, both quark distribution
functions and quark fragmentation functions.
In section III we present the diquark spectator approach and the
construction of the vertices. The results for the distribution and
fragmentation functions are given in section IV. 
We end with a summary and outlook.


\section{Quark correlation functions}

\subsection{Distribution functions}

In order to study the correlation function $\Phi$ it is useful to realize that
its form is constrained by the hermiticity properties of the fields and
invariance under parity operation.
The most general expression for $\Phi$ consistent with these
constraints is \cite{{ralston79},{piet95}}:
\bea \label{cf2}
\Phi(p,P,S) & = & M \,A_{1} +  A_{2}\, \sla P + A_{3} \,\sla p +
{A_{4} \over M} \,\sigma^{\mu \nu} P_{\mu} p_{\nu} + i \ A_5 \,p
\cdot S \ \ga  + M \,A_{6} \,\sla S \,\ga \nonumber \\ & &
+ {A_{7} \over M} \,p \cdot S \ \sla P \,\ga +
{A_{8} \over M} \,p \cdot S \ \sla p \,\ga +
i \,A_{9} \,\smn \,\ga \ S_{\mu} P_{\nu} \nonumber \\ & &
+ i \,A_{10} \,\smn \,\ga \,S_{\mu} p_{\nu} + 
i \,{A_{11} \over M^2} \,p \cdot S \ \smn \,\ga \,p_{\mu} P_{\nu}
+ {A_{12} \over M} \,\lc \,\gamma^{\mu} P^{\nu} p^{\rho} S^{\sigma},
\eea
where the amplitudes $A_i$ depend on $\sigma \equiv 2p\cdot P$ and
$\tau \equiv p^2$. Hermiticity requires all amplitudes $A_i (\sig, \tau)$ to
be real. Time reversal invariance can also be used and requires the amplitudes
$A_4$, $A_5$ and $A_{12}$ to be purely imaginary, hence they vanish.

In hard processes the hard momentum scale $q$ and the hadron momentum $P$ 
define the lightlike directions $n_\pm$. 
The momentum $P$ is parametrized as in Eq.~(\ref{parp}). The spin vector $S$ 
and the quark momentum $p$ are also expanded in the lightlike vectors
and transverse components:
\bea
&&p = {\xb (p^2 + \pts) \over x Q \sqrt{2}} \ n_-
+ {x Q \over \xb \sqrt{2}} \ n_+ + p_T, \\
&&S = - {\lam M \xb \over Q \sqrt{2}} \ n_- 
+ {\lam Q \over M \xb \sqrt{2}} \ n_+ + S_T. \label{sudakov1}
\eea
Thus $x$ represents the fraction of
the momentum in the $+$ direction carried by the quark inside the hadron.
In the transverse space the following projectors can be used
\bea
&&g_T^{\mu\nu} = g^{\mu\nu} - n_+^{\{\mu}n_-^{\nu\}},
\\
&&\epsilon_T^{\mu\nu} = \epsilon^{-+\mu\nu}.
\eea

Considering a hard scattering process up to order $1/Q^2$, the component
of $p$ along $n_-$ is irrelevant and one encounters the quantities
\be
\Phi^{[\Gam]} \left( x, \ptns \right) =
\left. \half \int dp^- \ \tr \left( \Phi \Gam \right) \right|_{p^+=xP^+, p_T}
=  \int [d\sigma d\tau\,\delta( \ )] \ \frac{\tr (\Phi \Gamma)}{4P^+},
\label{DiracProjPhi}
\ee
where we used the shorthand notation
\be
\left[ d\sigma d\tau \,\delta( \ )\right] = d\sigma d\tau \
\delta \left( \tau -x \sig + x^2M^2 + \pt \right).
\ee

The projections of $\Phi$ on different Dirac structures 
define distribution functions. They are related to integrals over linear
combinations of the amplitudes. The projections
\begin{eqnarray}
\label{tw2Phi}
\Phi^{[\gamma^+]}(x,\tv{p}) &\equiv&
f_1(x,\tv{p}^2)
\nnn & = &
\int[d\sigma d\tau\,\delta( \ )]
\left[A_2+xA_3\right],
\\
\nnn
\Phi^{[\gamma^+\gamma_5]}(x,\tv{p}) &\equiv&
\lambda g_{1L}(x,\tv{p}^2)+\frac{\ptinst}{M}g_{1T}(x,\tv{p}^2)
\nn \\ &=&
\int[d\sigma d\tau\,\delta( \ )] \Biggl\{
\lambda \left[ -A_6
-\left(\frac{\sigma - 2x\,M^2}{2M^2}\right)\;\left(A_7+xA_8\right)\right]
+\frac{\ptinst}{M}\;\left(A_7+xA_8\right)\Biggr\}, \\
\nnn
\Phi^{[i\sigma^{i+}\gamma_5]}(x,\tv{p}) &\equiv&
S_T^i h_{1T}(x,\tv{p}^2)
+\frac{p_T^i}{M}\left(\lambda h_{1L}^\perp(x,\tv{p}^2)
+\frac{\ptinst}{M}h_{1T}^\perp(x,\tv{p}^2)\right)
\nn \\ &=&
\int[d\sigma d\tau\,\delta( \ )] \Biggl\{
-S_T^i (A_9+xA_{10})
+\frac{\lambda\,p_T^i}{M}\left[A_{10}-
\left(\frac{\sigma - 2x\,M^2}{2M^2}\right)\;A_{11}\right]
\nn \\ && \qquad\qquad\qquad \mbox{}
+\frac{p_T^i}{M}\;\frac{\ptinst}{M}\;A_{11} \Biggr\},
\end{eqnarray}

\nin are leading in $1/Q$. For the distribution functions, this is indicated by 
the subscript $1$ in the names of the functions \cite{piet96}. 
The following projections occur with a
pre-factor $M/P^+$, which signals the subleading (or higher twist) 
nature of the corresponding distribution functions
\begin{eqnarray}
\label{tw3Phi}
\Phi^{[1]}(x,\tv{p}) &\equiv&
\frac{M}{P^+}\,e(x,\tv{p}^2)
\nn \\ &=&
\frac{M}{P^+}\int[d\sigma d\tau\,\delta( \ )] \ A_1,
\\
\nnn
\Phi^{[\gamma^i]}(x,\tv{p}) &\equiv&
\frac{p_T^i}{P^+}\,f^\perp(x,\tv{p}^2)
\nn \\ &=&
\frac{M}{P^+}\int[d\sigma d\tau\,\delta( \ )] \frac{p_T^i}{M}\;A_3, \\
\nnn
\Phi^{[\gamma^i\gamma_5]}(x,\tv{p}) &\equiv&
\frac{M\,S_T^i}{P^+} \,g^\prime_T(x,\tv{p}^2)
+\frac{p_T^i}{P^+}\left(\lambda g_L^\perp(x,\tv{p}^2)
+\frac{\ptinst}{M}g_T^\perp(x,\tv{p}^2)\right)
\nn \\ &=&
\frac{M}{P^+}\int[d\sigma d\tau\,\delta( \ )] \left\{
-S_T^i\;A_6
-\frac{\lambda\,p_T^i}{M}\,\left(\frac{\sigma - 2x\,M^2}{2M^2}\right)\;A_8
+ \frac{p_T^i}{M}\,\frac{\ptinst}{M}\;A_8 \right\}, \\
\nnn
\Phi^{[i\sigma^{ij}\gamma_5]}(x,\tv{p}) &\equiv&
\frac{S_T^ip_T^j-S_T^jp_T^i}{P^+}\,h_T^\perp(x,\tv{p}^2)
\nn \\ &=&
\frac{M}{P^+}\int[d\sigma d\tau\,\delta( \ )]
\frac{S_T^ip_T^j-S_T^jp_T^i}{M}\;\left[ -A_{10}\right] ,
\\
\nnn
\Phi^{[i\sigma^{+-}]}(x,\tv{p}) &\equiv&
\frac{M}{P^+}\left(\lambda h_L(x,\tv{p}^2)
+\frac{\ptinst}{M} \,h_T(x,\tv{p}^2)\right)
\nn \\ &=&
\frac{M}{P^+}\int[d\sigma d\tau\,\delta( \ )] \Biggl\{
\lambda\,\Biggl[-A_9-\frac{\sigma}{2M^2}\,A_{10}
+\left(\frac{\sigma - 2x\,M^2}{2M^2}\right)^2\,A_{11}\Biggr]
\nn \\ && \qquad\qquad\qquad \mbox{}
- \frac{\ptinst}{M}\,
\left(\frac{\sigma - 2x\,M^2}{2M^2}\right)\,A_{11}
\Biggr\}.
\end{eqnarray}

\begin{figure}[ht]
\begin{center}
\epsfig{file=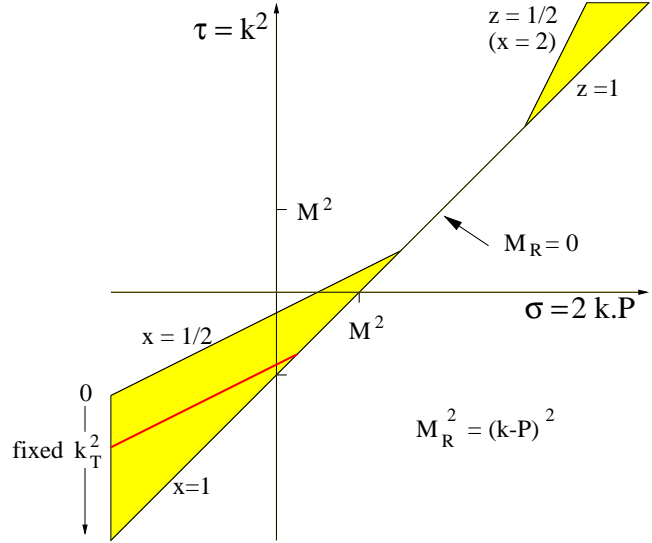,width=0.5\textwidth,angle=0}
\caption{\label{support} The $\delta$-function constraint in the 
$\sigma$-$\tau$ plane (using quark momentum $k$ and hadron momentum $P$)
coming from fixing $x$ and $\bm k_T^2$ in the
expression for the distribution functions $F(x,\bm k_T^2)$ 
(and similarly for the fragmentation functions $D(z,z^2 \bm k_T^2)$) 
and the full integration regions for the $\bm k_T$ integrated functions 
$F(x)$ (and similarly for $D(z)$). The latter region is determined by
$\bm k_T^2 \ge 0$ and $M_R^2 \ge 0$. }
\end{center}
\end{figure}

The constraint of the $\delta$-function in the integration over
$\sig$ and $\tau$ is indicated in Fig.~\ref{support}. 
Furthermore, the integration is restricted 
to the region $M_R^2 \equiv (P-p)^2 \ge 0$. This leads to the vanishing of the 
distribution functions at $x = 1$. 

If a generic distribution function is written as
\be \label{integrated1}
F(x,\pts) = \int[d\sigma d\tau\,\delta( \ )] 
\ G\left(A_i(\sig,\tau),\sig,x\right),
\ee
symmetric integration over ${\bm p}_T$ gives
\be 
F(x) = \int[d\sigma d\tau \theta( \ )] \ G\left(A_i(\sig,\tau),\sig,x\right),
\ee
where
\be
[d\sig d\tau \,\theta( \ )] 
= d\sig d\tau \ \theta \left( x\sig  -x^2M^2 -\tau \right).
\ee
The region covered by the $\theta$-function (for $x= 1/2$)
is the lower shaded region 
in Fig.~\ref{support} corresponding to ${\bm p}_T^2 \ge 0$.
Only the terms involving the distribution functions $f_1$, $g_1=g_{1L}$,
$h_1=h_{1T}+({\bm p}_T^2/2M^2)\,h_{1T}^\perp$, $e$, $g_T = g^\prime_T +
({\bm p}_T^2/2M^2)\,g_T^\perp$ and $h_L$ are non-vanishing upon 
integration over ${\bm p}_T$. The integrated functions $\fox$, $\gox$ and
$\hox$ have the well known probabilistic interpretations.
$f_1(x)$ gives the probability of finding a quark with light-cone 
momentum fraction $x$ in the $+$ direction (and any transverse momentum).
$g_1(x)$ is a chirality distribution: in a hadron that is in a positive
helicity eigenstate ($\lambda = 1$), it measures the probability of finding a 
right-handed quark with light-cone momentum fraction $x$ minus the
the probability of finding a left-handed quark with the same light-cone
momentum fraction (and any transverse momentum). $h_1(x)$ is a transverse
spin distribution: in a transversely polarized hadron, it measures the
probability of finding quarks
with light-cone momentum fraction $x$ polarized along the direction of
the polarization of the hadron minus the probability of finding quarks
with the same light-cone momentum fraction polarized along the direction
opposite to the polarization of the hadron. The twist three functions have
no intuitive partonic interpretation. Nevertheless, they are well defined as
hadronic matrix elements via Eqs.~(\ref{qcf1}) and (\ref{qcf2}), and
their projections.

We note the appearance of higher $\bm p_T^2$-moments,
\bea 
F^{(n)}(x) &\equiv& \int d^2 \ptns \ \left( {\pts \over 2M^2} \right)^n
F(x,\pts)
\nnn
&=&
\pi \int [d\sig d\tau \,\theta( \ )] 
\ \left(\frac{x \sig - x^2M^2 - \tau}{2M^2}\right)^n 
\ G\left(A_i(\sig,\tau),\sig,x\right), 
\label{integrated2}
\eea
such as $h_{1T}^{\perp (1)}$ and $g_T^{\perp (1)}$.
The equality in Eq.~(\ref{integrated2}) is obtained using
the azimuthal symmetry of the distribution functions, which depend only on $x$
and ${\bm p}_T^2$.
In the weighted integration, $\int d^2\bm p_T\,p_T^i \ldots$ one will encounter 
the functions $g_{1T}^{(1)}$ and $h_{1L}^{\perp (1)}$.

The distribution functions cannot be all independent because their number is
larger than the number of amplitudes $A_i$. This is reflected in relations
such as 
\bea \label{piet1}
\gtx & = & \gox + {d \over dx} \ \gotox, \\
\hlx & = & \hox - {d \over dx} \ \holperpox, \label{piet2} \\
\htox & = & - \frac{1}{2}\,\frac{d}{dx} \ \hotperptx,
\eea
\nin which can be obtained using their 
explicit expressions in terms of the amplitudes.

The functions $\gtwo = \gt - \go $ and $\htwo = 2(\hl - \ho) $
thus satisfy the sum rules
\bea \label{bcsr}
\int_0^1 dx \ \gtwox & = & - \goto (0), \\
\int_0^1 dx \ \htwox & = & 2\,\holperpo (0), \label{bsr}
\eea
which are a direct consequence of (\ref{piet1}) and (\ref{piet2}).
If the functions $\goto$ and $\holperpo$ vanish at the origin, we
rediscover the Burkhardt-Cottingham
sum rule \cite{bc70} and the Burkardt sum rule \cite{burkardt93}.
These sum rules (Eqs.~(\ref{bcsr}) and (\ref{bsr}) with vanishing right-hand
sides) can also be derived using Lorentz covariance for the
expectation values of local operators \cite{burkardt95}.
In our approach this would imply constraints on the amplitudes $A_i$.


\subsection{Fragmentation functions}

The correlation function $\Delta$ is also constrained by the hermiticity
properties of the fields and invariance under parity operation, leading
to an expansion identical to that in Eq.~(\ref{cf2}) with the replacements
$\left\{p,P,S,M\right\}\to
\left\{k,P_h,S_h,M_h\right\}$ \cite{piet96}
and with real amplitudes, say $B_i$, now depending on
$\tau_h \equiv k^2$ and $\sigma_h \equiv 2k\cdot P_h$. Time reversal
invariance does not imply any constraints on the amplitudes, thus
$B_4$, $B_5$ and $B_{12}$, referred to as `T-odd', 
are, in general, non-vanishing.
(For a discussion on T-odd fragmentation functions, see \cite{rainer97}).

In hard processes one encounters the quantities
\be \label{DiracProjDelta}
\Del^{[\Gam]} \left( z, \ktns \right) = \left. {1 \over 4z} \
\int dk^+ \ \tr \left( \Del \Gam \right) \right|_{k^-=P_h^-/z, k_T}
= \int \left[ d\sig_h d\tau_h \,\delta( \ )\right]
\ \frac{\tr (\Del \Gam )}{8z\,P_h^-},
\ee
where we used the shorthand notation
\be
\left[ d \sig_h d \tau_h \,\delta( \ )\right] =
d\sig_h d\tau_h \ \del \left( \tau_h - \frac{\sig_h}{z} + \frac{M_h^2}{z^2} 
+ \kt \right).
\ee
The momentum $P_h$ is parametrized as in Eq.~(\ref{parph})
and is used to define the lightlike vectors, in terms of which 
the vectors $S_h$ and $k$ can also expanded:
\bea
&& k = {z_h Q \over z \sqrt{2}} \ n_-
+ {z (k^2 + \kt) \over z_h Q \sqrt{2}} \ n_+
+ k_T,
\label{kexps}
\\
&&S_h = {\lam_h z_h Q \over \mh \sqrt{2}} \ n_- -
{\lam_h \mh \over z_h Q \sqrt{2}} \ n_+ + S_{hT}. \label{sudakov2}
\eea
Thus $z$ is the fraction of the momentum in the $-$ direction carried by
the hadron $h$ originating from the fragmentation of the quark. The spin
vector satisfies $P_h\cdot S_h$ = 0 and for a pure state
$-S_h^2$ = $\lambda_h^2 + \bm S_{hT}^2$ = 1. 

Comparing the above equations with the case of the distribution functions,
one sees that the relations between the Dirac projections
$2z\,\Delta^{[\Gamma]}(z,\bm k_T)$ and the amplitudes 
are identical to those for $\Phi^{[\Gamma]}(x,\bm p_T)$ after the replacements
$\left\{x,\sigma,\tau,\tv{p},P,S_T,\lambda,M,A_i,
\pm-\mbox{components}\right\}\to
\left\{1/z,\sigma_h,\tau_h,\tv{k},P_h,S_{hT},\lambda_h,M_h,B_i,
\mp-\mbox{components}\right\}$,
except for additional parts originating from the T-odd functions.
Furthermore, the definition of fragmentation functions follow the general
procedure used to define distribution functions.
We use for the names of the fragmentation functions capital letters
(with the only exception for the counterparts of $f_{..}$ functions which are 
called $D_{..}$). For example,
\begin{eqnarray}
\Delta^{[\gamma^-]}(z,\tv{k}) &\equiv&
D_1(z,\bm k_T^{\prime 2})
+\frac{\epsilon_{T}^{ij}k_{Ti}S_{hTj}}{M_h} D_{1T}^\perp(z,\bm k_T^{\prime 2})
\nnn
&=& \frac{1}{2z}\int[d\sigma_h d\tau_h\,\delta( \ )]
\Biggl\{ \left[B_2 + \frac{1}{z}\,B_3\right]
+ \frac{\epsilon_{T}^{ij}k_{Ti}S_{hTj}}{M_h}\,B_{12}\Biggr\},
\end{eqnarray}
where $\kpns = -z\ktns$. The choice of arguments $z$ and $\kpns$ in the
fragmentation functions is worth a comment. In the expansion of $k$ in
Eq.~(\ref{kexps}) the quantities $1/z$ and $\ktns$ appear in a natural way. 
However, in the interpretation of $\Delta$ as a decay function of quarks,
the variable $z$ as the ratio of $P_h^-/k^-$ is more adequate.
Applying a Lorentz transformation that leaves the $-$ component (and hence
the definition of $z$) unchanged, one finds that $\kpns = -z \ktns$ is
the transverse component of hadron $h$ with respect to the quark momentum.

Further, we only display the additional parts of projections which come from 
the time reversal odd amplitudes and which, thus, have no counterparts
in the distribution functions. 
In the leading twist projections there is only one other projection with
an additional term,
\begin{eqnarray}
\Delta^{[i\sigma^{i-}\gamma_5]}(z,\tv{k}) &\equiv&
\ldots+\frac{\epsilon_T^{ij}k_{Tj}}{M_h} H_1^\perp(z,\tv{k}^{\prime 2})
\nnn
& = & \ldots +\frac{1}{2z}\int[d\sigma_h d\tau_h\,\delta( \ )]
\,\frac{\epsilon_T^{ij}k_{Tj}}{M_h}
\left[-B_4\right].
\end{eqnarray}
At subleading twist there are five additional T-odd structures:
\begin{eqnarray}
\Delta^{[\gamma^i]}(z,\tv{k}) &\equiv&
\ldots+\frac{M_h}{P_h^-}\left(
\lambda_h\frac{\epsilon_T^{ij}k_{Tj}}{M_h}D_L^\perp(z,\tv{k}^{\prime 2})
+\epsilon_T^{ij}S_{hTj}D_T(z,\tv{k}^{\prime 2})\right)
\nnn 
&=&
\ldots +\frac{M_h}{2zP_h^-}\int[d\sigma_h d\tau_h\,\delta( \ )] \left\{
-\lambda_h\frac{\epsilon_T^{ij}k_{Tj}}{M_h}\,B_{12}
-\epsilon_T^{ij}S_{hTj}\left(\frac{\sigma_h-2M_h^2/z}{2M_h^2}\right)
B_{12} \right\},
\\
\nnn
\Delta^{[i\gamma_5]}(z,\tv{k}) &\equiv&
\ldots+\frac{M_h}{P_h^-}\left(\lambda_h E_L(z,\tv{k}^{\prime 2})
+\frac{\ktinsht}{M_h}E_T(z,\tv{k}^{\prime 2})\right)
\nnn 
&=&
\ldots+\frac{M_h}{2zP_h^-}\int[d\sigma_h d\tau_h\,\delta( \ )] \left\{
-\lambda_h \left(\frac{\sigma_h-2M_h^2/z}{2M_h^2}\right)\,B_5
+\frac{\ktinsht}{M_h}\, B_5 \right\},
\\
\nnn
\Delta^{[i\sigma^{ij}\gamma_5]}(z,\tv{k}) &\equiv&
\ldots+\frac{M_h}{P_h^-}\epsilon_T^{ij}H(z,\tv{k}^{\prime 2})
\nnn
& =&
\ldots + \frac{M_h}{2zP_h^-}\int[d\sigma_h d\tau_h\,\delta( \ )] \left\{
\epsilon_T^{ij}\left(\frac{\sigma_h-2M_h^2/z}{2M_h^2}\right)\,B_4
\right\}.
\end{eqnarray}

The constraint imposed by the $\delta$-function in the $\sigma_h$-$\tau_h$
plane is also indicated in Fig.~\ref{support}. The integration is restricted to
the region $M_R^2 = (P_h-k)^2 \ge 0$, which implies that the 
fragmentation functions vanish
at $z = 1$. We note the reciprocity of $x$ and
$z$, i.e., the constraint for $z = 1/2$ is the same as one would have for
$x$ = 2. Note, however, that the integration involves different regions.
For the distributions one has (roughly) spacelike quark momenta, for the 
fragmentation timelike quark momenta.
If a generic quark fragmentation function is given by
\be
D(z, \kp) = {1 \over 2z} \int [d\sig_h d\tau_h\,\delta( \ )] \ 
G(B_i(\sig_h, \tau_h),\sig_h,z),
\ee
the integrated functions are given by
\bea 
D^{(n)}(z) & \equiv & z^2\int d^2\ktns \left( {\kt \over 2\mh^2} \right)^n
D(z, \kt) 
\nnn
& = & \frac{\pi\,z}{2} \int [d\sig_h d\tau_h\,\theta( \ )]
\left(\frac{\sigma_h-2M_h^2/z}{2M_h^2}\right)^n G(B_i(\sig_h,\tau_h), \sig_h,z),
\label{integrated4}
\eea
where
\be
[d\sig_h d\tau_h\,\theta( \ )] = d\sig_h d\tau_h \ \theta
\left( \frac{\sig_h}{z} - \frac{M_h^2}{z^2} - \tau_h\right).
\ee
Non-vanishing upon integration over $\tv{k}$ are the fragmentation functions
$D_1$, $G_1 = G_{1L}$, $H_1 = H_{1T}+(\tv{k}^2/2M_h^2)\,H_{1T}^\perp)$, $E$,
$G_T = G_T^\prime + (\tv{k}^2/2M_h^2)\,G_T^\perp$, $H_L$ and $D_T$.

As for the distributions, the integrated fragmentation functions are not
all independent. Using Eq.~(\ref{integrated4}) one obtains relations such 
as
\bea
\Elz & = & z^3 \ {d \over dz} \left[ \frac{\Etoz}{z} \right], \\
\Dtz & = & z^3 \ {d \over dz} \left[ \frac{\Dotperpoz}{z} \right], \\
\Gtz & = & \Goz - z^3 \ {d \over dz} \left[ \frac{\Gotoz}{z} \right], \\
\Hlz & = & \Hoz + z^3 \ {d \over dz} \left[ \frac{\Holperpoz}{z} \right], \\
\Hz & = & z^3 \ {d \over dz} \left[ \frac{\Hoperpoz}{z} \right],
\eea
leading to
\be
\int_0^1 dz \ {\Elz \over z^3} = \lim_{z \rightarrow 0} 
\frac{E_T^{(1)}(z)}{z},
\ee
and similar ones for $D_T$, $\Gtwo = \Gt - \Go$, $\Htwo = 2(\Hl - \Ho)$ and
$H$. Provided that the functions labelled  with superscript $(1)$ vanish at
the origin faster than one power of $z$, the right hand side vanishes.
Finally, let us remark that this formalism can be easily extended to
include antiquarks \cite{piet96}.


\section{The spectator model}

\begin{figure}[t]
\begin{center}
\epsfig{file=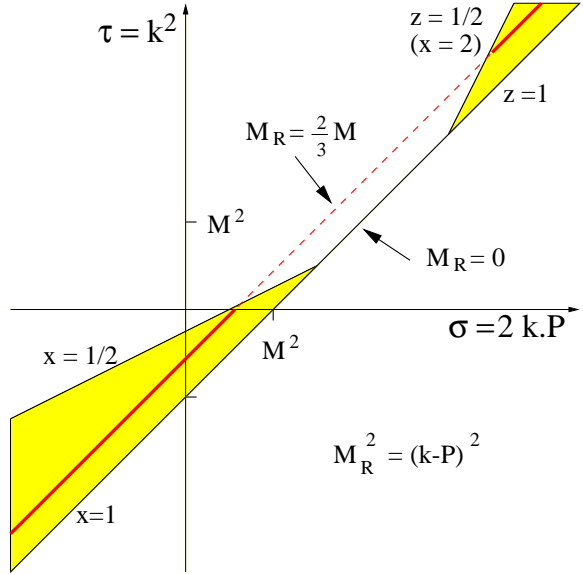,width=0.45\textwidth,angle=0}
\caption{\label{support_spectator} \protect
The constraint in the $\sigma-\tau$ plane
coming from fixing the spectator mass $M_R$ (compare with Fig.~\ref{support}).}
\end{center}
\end{figure}

The basic idea of the
spectator model is to treat the intermediate states that can be inserted
in the definition of the correlation function $\Phi$ in Eq.~(\ref{qcf1}), or
which are explicitly displayed in the definition of the correlation function
$\Delta$ in Eq.~(\ref{qcf2}), as a state with a definite mass. In other words,
we make a specific ansatz for the spectral decomposition of these
correlation functions. This may be best illustrated using the support plot
in $\sigma$ and $\tau$. In this plot the mass $M_R$ of the remainder, called the
spectator, is constant along the lines $(P-k)^2 = \tau - \sigma + M^2 = M_R^2$,
as indicated in Fig.~\ref{support_spectator}. 
The quantum numbers of the intermediate state are those determined by 
the action of the quark field on the state $\vert P,S\rangle$, 
hence the name {\em diquark spectator}. 
In the most naive picture of the quark structure of the nucleon, such that 
in its rest frame all quarks are in $1/2^+$ orbitals, 
the spin of the diquark system can be either 0 (scalar diquark $s$) or 1 (axial
vector diquark $a$). 
For a pion state we have an {\em antiquark spectator}. 
The inclusion of antiquark and gluon distributions requires
a more complex spectral decomposition of intermediate states. 
Here, we restrict ourselves to the simplest case.
The correlation function $\Phi$ (the correlation function $\Del$ 
will be treated later) is then given in the spectator model by
\begin{equation}
\label{qcfdq1}
\Phi^{R}_{ij}(p,P,S)=
\frac{1}{(2 \pi)^3} \;
\langle P,S | \bpsi_j(0) | X^{(\lambda)} \rangle \;
\theta ( P_R^+ )\;\delta \left[ (p-P)^2 - M_{R}^2 \right] \;
\langle X^{(\lambda)} | \psi_i(0) | P,S \rangle,
\end{equation}
where $P_R$ = $P-p$ and $X^{(\lambda)}$ represents the spectator and its
possible spin states (indicated with $\lambda$). We will project onto
different spins in the intermediate state and allow for different
spectator masses.  

We start with the correlation function $\Phi$ for a nucleon. 
The matrix element appearing in the RHS of (\ref{qcfdq1}) is given by
\begin{equation}
\langle X_s | \psi_i(0) | P, S \rangle =
{\left( {i \over \sla p -m } \right)}_{ik} \;\Upsilon^s_{kl} \ U_l(P,S),
\end{equation}
in the case of a scalar diquark, or by
\begin{equation}
\langle X_a^{(\lam)} | \psi_i(0) | P, S \rangle =
\eps_\mu^{*(\lam)}
{\left( {i \over \sla p -m } \right)}_{ik} \;\Upsilon^{a\mu}_{kl} \ U_l(P,S),
\end{equation}
for a vector diquark. The matrix elements consist of a nucleon-quark-diquark 
vertex $\Upsilon (N)$ yet to be specified, the Dirac spinor
for the nucleon $U_l(P,S)$, a quark propagator for the untruncated quark line
($m$ is the constituent mass of the quark) 
and a polarization vector $\eps_\mu^{*(\lam)}$
in the case of an axial vector diquark. The next step is to fix the Dirac
structure of the nucleon-quark-diquark vertex $\Upsilon$. We assume the
following structures:
\bea
&&\Upsilon^s (N) = {\bf 1} \ g_s(p^2),
\\ &&
\Upsilon^{a \mu} (N) = {g_a(p^2) \over \sqrt{3}}
\,\gamma_\nu\gamma_5\,\frac{\sla P + M}{2M}
\,\left( - g^{\mu\nu} + \frac{P^\mu P^\nu}{M^2}\right)
=\frac{g_a(p^2)}{\sqrt{3}}\,\gamma_5\,\left(\gamma^\mu + \frac{P^\mu}{M}\right).
\eea
The functions 
$\gr (p^2)$ (where $R$ is $s$ or $a$) are form factors that take into 
account the composite structure of the nucleon and the diquark spectator.
In the choice of vertices, the factors and projection operators are
chosen to assure that in the target rest frame, 
where the nucleon spinors have only upper components,
the diquark spin 1 states are purely spatial and in
which case the axial vector diquark vertex reduces to $\chi_N^\dagger
\bm \sigma \cdot \bm \epsilon \chi_q$.
The most general structure of the vertices can be found in \cite{thomas94}.
With our choices, we find
\begin{equation}
\label{qcfdq2}
\Phi^R(p, P, S) =
{\vert \gr (p^2)\vert^2 \over 2(2 \pi)^3} \
{\del \left(\tau-\sigma+ M^2 -M_R^2 \right)
\over {\left( p^2 -m^2 \right)}^2} \,(\sla p + m)\,\left(\sla P + M \right)
\,(1 + \ar \ga \sla S) \ (\sla p +m),
\end{equation}
where $\ar$ is a spin factor, which takes the values 
$a_s=1$ and $a_a=-1/3$. In obtaining this result we used as the
polarization sum for the axial vector diquark in the form
$\sum_\lambda\eps_\mu^{*(\lam)} \eps_\nu^{(\lam)}=-g_{\mu\nu}+P_\mu P_\nu/M^2$,
which is consistent with the choice that the axial vector diquark spin states
are purely spatial in the nucleon rest frame.
We will use the same form factors for scalar and axial vector diquark:
\be
\label{ff}
g(\tau) = N\,\frac{\tau - m^2}{\left| \tau - \Lam^2 \right|^\alpha}.
\ee
The quantity $\Lam$ is another parameter of the model which ensures that 
the vertex is cut off if the virtuality of the quark leg is much larger
than $\Lam^2$. $N$ is a normalization constant. 
This choice of form factor has the
advantage of killing the pole of the quark propagator as suggested in
\cite{thomas94}.

In the same way, one can write down a simple spectator 
model for the pion. The matrix element can be written as
\begin{equation}
\langle X^{(\alp)} | \psi_i(0) | P_\pi \rangle =
{\left( {i \over \sla p -m } \right)}_{ik} \;\Upsilon_{kl} \ v_l^{(\alp)}.
\end{equation}
The spinor $v_l^{(\alp)}$ describes the spin state of the antiquark spectator. 
The simplest vertex is given by
\be
\Upsilon (\pi) = \frac{g(p^2)}{\sqrt{2}}\,\frac{\sla P_\pi + 
M_\pi}{2M_\pi}\,\gamma_5.
\ee
Taking for the spectator antiquark spin sum $\sum_\alp v_l^{(\alp)}
\bar v_l^{(\alp)}$ = $\sla P_\pi - M_\pi$ one arrives at precisely the same
expression as for the nucleon (Eq.~(\ref{qcfdq2})) with $\ar = 0$. 

{}From the correlation function $\Phi$ one easily obtains the distribution
functions. Taking out the explicit $\delta$-function,
\be
\Phi^R(p,P,S)
= \tilde \Phi(p,P,S) \ \delta\left( \tau-\sigma+M^2-M_R^2\right),
\ee
one finds immediately from Eq.~(\ref{DiracProjPhi}) the result
\be
\Phi^{[\Gamma]}(x,\bm p_T)
= \left. \frac{\tr (\tilde \Phi \Gamma)}{4(1-x)P^+}
\right|_{\normalsize \tau = p^2(x,\bm p_T^2)},
\ee
with
\be
-p^2(x,\bm p_T^2) = \frac{\bm p_T^2}{1-x} + \frac{x}{1-x}\,M_R^2 - x\,M^2.
\ee

We now turn to the fragmentation functions. The calculation is very similar
to the case of the distribution functions, involving the same type of
matrix elements. Further, we assume
that the hadron $h$ has no interactions with the the spectator.
This allows us to use a free spinor to describe this outgoing hadron. 
Then we see that the correlation function $\Delta$ is the same as the
one needed for the distributions, after obvious replacements in the 
arguments, namely
\be
\label{frag}
\Delta^R(k, P_h, S_h) =
{\vert \gr (k^2)\vert^2 \over 2(2 \pi)^3} \
{\del \left(\tau_h-\sigma_h+ M_h^2 -M_R^2 \right)
\over {\left( k^2 -m^2 \right)}^2}\,(\sla k + m)\,\left(\sla P_h + M_h \right)
\,(1 + \ar \ga \sla S_h) \,(\sla k +m).
\ee
A direct consequence is 
\be \label{reciprocity}
\Delta^{[\Gamma]}(z,\bm k_T) = \frac{1}{2z}\,\Phi^{[\Gamma^\prime]}\left(
\frac{1}{z},\bm k_T\right) = \frac{1}{2z}\,\Phi^{[\Gamma^\prime]}\left(
\frac{1}{z},-\,\frac{\bm k_T^\prime}{z}\right),
\ee
where $\Gamma^\prime$ and $\Gamma$ involve an interchange of $+$ and $-$ 
components. Writing $\Delta(k,P_h,S_h)$ = 
$\tilde \Delta(k,P_h,S_h)\,\delta \left((k-P_h)^2 - M_R^2\right)$,
Eq.~(\ref{DiracProjDelta}) leads to
\be
\Delta^{[\Gamma]}(z,\bm k_T) =
\left. \frac{\tr (\tilde \Delta \Gamma)}{8(1-z)P_h^-}
\right|_{\normalsize \tau_h = k^2(z,\bm k_T^2)},
\ee
with
\be
k^2(z,\bm k_T^2) = \frac{z}{1-z}\,\bm k_T^2 + \frac{M_R^2}{1-z} + 
\frac{M_h^2}{z}.
\ee
The consequence of using free spinors to describe the outgoing hadron is 
that all T-odd fragmentation functions vanish and
we have a one-to-one correspondence between distribution and fragmentation
functions. As can be seen in Fig.~\ref{support_spectator} the actual 
behavior of the distribution and fragmentation functions comes from 
different regions in $\tau$, roughly spacelike and timelike, respectively. 
Therefore, the above reciprocity (Eq.~(\ref{reciprocity}))
is of use for the analytic expressions, less for the actual values.

\newpage

\section{Results and discussion}

\subsection{Distribution functions of the nucleon}

Using the expression in Eq.~(\ref{qcfdq2}) we can
compute the amplitudes $A_i$ shown in Eq.~(\ref{cf2}).
Taking out some common factors by defining
\be
A_i = {N^2 \over 2(2\pi)^3} \
{\del \left( \tau - \sig +M^2 - M_R^2 \right) \over
\left| \tau - \Lam^2 \right|^{2\alpha}} \ \tilde{A}_i,
\ee
we obtain, as expected, the T-odd amplitudes 
$\tilde A_4$ = $\tilde A_5$ = $\tilde A_{12}$ = 0, and
\bea \label{a1}
\tilde{A}_1 & = &
{m \over M}\,\left( (M+m)^2 - M_R^2\right) +
(\tau - m^2) \left( 1 + {m \over M} \right) , \\
\tilde{A}_2 & = & - \left( \tau - m^2 \right), \\
\tilde{A}_3 & = & (M+m)^2 - M_R^2 + (\tau-m^2) , \\
\tilde{A}_6 & = &
- \ar \,\left[{m \over M}\,\left( (M+m)^2 - M_R^2 \right)
+ (\tau-m^2) \left( 1 + {m \over M} \right)\right], \\
\tilde{A}_7 & = & 2\, \ar \, mM, \\
\tilde{A}_8 & = & 2\, \ar \,M^2 , \\
\tilde{A}_9 & = & \ar \,(\tau - m^2), \\
\tilde{A}_{10} & = &
- \ar \,\left[(M+m)^2 - M_R^2 + (\tau-m^2) \right], \\
\tilde{A}_{11} & = & -2\,a_R\,M^2. \label{a11}
\eea
Introducing the function $\lamR^2(x)$ such that
\be
\Lambda^2 - p^2 = \frac{\bm p_T^2 + \lamR^2(x)}{1-x},
\ee
which implies
\be
\lamR^2(x) = \Lambda^2(1-x) + xM_R^2 - x(1-x)M^2,
\ee
one gets the following results for the distribution functions,
\bea
\foxp & = & {N^2 \,(1-x)^{2\alpha - 1} \over 16 \pi^3} \ 
{ {\left( xM + m \right)}^2 + \pts \over
\left( \pts + \lamR^2 \right)^{2\alpha}}, \\
\golxp & = & \ar \,{N^2 \,(1-x)^{2\alpha - 1} \over 16 \pi^3} \ 
{ {\left( xM + m \right)}^2 - \pts \over
\left( \pts + \lamR^2 \right)^{2\alpha}}, \\
\gotxp & = & \ar \,{N^2\ \,(1-x)^{2\alpha - 1} \over 8 \pi^3} \ 
{ M(xM + m) \over
\left( \pts + \lamR^2 \right)^{2\alpha}}, \\
\hotxp & = & \ar \,\foxp, \\
&& \nn \\
\holperpxp & = & -\gotxp,\\
&& \nn \\
\hotperpxp & = & - \, \ar \, {N^2 \,(1-x)^{2\alpha - 1} \over 8 \pi^3} \ 
{ M^2 \over \left( \pts + \lamR^2 \right)^{2\alpha}}, \\
\expp & = & {N^2 \,(1-x)^{2\alpha - 2} \over 16 \pi^3} \
{(1-x) (xM + m ) (M + m) - M_R^2 \left( x + {m \over M} \right) 
- \left( 1 + {m \over M} \right) \pts \over
\left( \pts + \lamR^2 \right)^{2\alpha}}, \\
\fperpxp & = & {N^2 \,(1-x)^{2\alpha - 2} \over 16 \pi^3} \
{(1-x^2) M^2 + 2mM (1-x) - M_R^2 - \pts \over
\left( \pts + \lamR^2 \right)^{2\alpha}}, \\
\gprimetxp & = & \ar \,\expp,\\
&& \nn \\
\glperpxp & = & - \, \ar \, {N^2 \,(1-x)^{2\alpha - 2} \over 16 \pi^3} \
{(1 - x)^2M^2 - M_R^2 - \pts \over
\left( \pts + \lamR^2 \right)^{2\alpha}}, \\
\gtperpxp & = & \ar \,{N^2 \,(1-x)^{2\alpha - 1} \over 8 \pi^3} \
{ M^2 \over \left( \pts + \lamR^2 \right)^{2\alpha}}, \\
\htperpxp & = & \ar \, \fperpxp,\\
&& \nn \\
\hlxp & = & \ar \,{N^2 \,(1-x)^{2\alpha - 2} \over 16 \pi^3} \
{ (1-x) (xM+m) (M + m) 
- \left( x + {m\over M} \right) M_R^2
+ \left( 1 - 2x - {m \over M} \right) \pts \over
\left( \pts + \lamR^2 \right)^{2\alpha}},  \\
\htxp & = & -\glperpxp.
\eea
Although there is a certain freedom in the choice of the
parameters, one immediately sees that the occurence of singularities in
the integration region (see Fig.~\ref{support_spectator}) will cause 
problems which are avoided
if there is no zero in the denominator. The requirement that 
$\lamR^2(x)$ is positive implies for the distribution functions ($0 \le 
x \le 1$) 
\be
\label{condition1}
M_R > M-\Lam,
\ee
while for the fragmentation functions (using reciprocity, we have to 
look at $x \ge 1$) it leads to
\be
\label{condition2}
M_R >\Lam - M_h.
\ee
Provided condition (\ref{condition1}) is fulfilled, one obtains the integrated 
distribution functions,
\bea
\fox & = & {N^2 \,(1-x)^{2 \alp -1} \over 32 \pi^2\,(\alpha-1)(2\alpha-1)} 
\ {2(\alpha-1)\,( xM + m)^2 + \lamR^2(x) \over 
\left(\lamR^2(x)\right)^{2\alpha - 1}}, \\
\gox & = & {N^2 \ar \,(1-x)^{2 \alp -1} \over 32 \pi^2\,(\alpha-1)(2\alpha-1)} 
\ {2(\alpha-1)\,( xM + m)^2 - \lamR^2(x) \over 
\left(\lamR^2(x)\right)^{2\alpha - 1}}, \\
\hox & = & {N^2 \ar \,(1-x)^{2 \alp -1} \over 16 \pi^2 (2 \alp -1)} \
{(xM + m)^2 \over \left(\lamR^2(x)\right)^{2\alpha - 1}}, \\
e(x) & = & {N^2 \,(1-x)^{2 \alp -2} \over 32 \pi^2 (\alp -1)(2\alp-1)} \
{2( \alp -1) \left( x + {m \over M} \right) \left[ (1-x) (M + m)M 
- M_R^2 \right] - 
\left( 1 + {m \over M} \right) \lamR^2(x) 
\over \left(\lamR^2(x)\right)^{2\alpha - 1}}, \\
\gtx & = & {N^2 \ar \,(1-x)^{2 \alp -2} \over 32 \pi^2 (\alp -1)(2\alp-1)} \
{2( \alp -1) \left( x + {m \over M} \right) \left[ (1-x) (M + m)M 
- M_R^2 \right] - \left( x + {m \over M} \right) \lamR^2(x) 
\over \left(\lamR^2(x)\right)^{2\alpha - 1}}, \\
\hlx & = & {N^2 \ar \,(1-x)^{2 \alp -2} \over 32 \pi^2 (\alp -1) (2 \alp -1)} \
{ 2( \alp -1) \left( x+ {m \over M} \right) 
\left[ (1-x) \left( M + m \right)M - M_R^2 \right]
+ \left( 1 - 2x - {m \over M} \right) \lamR^2(x) 
\over \left( \lamR^2(x) \right)^{2\alp-1} }.
\eea
Examples of the $\bm p_T^2/2M^2$-weighted distributions are
\be
\gotox = - \holperpox = {N^2 \ar \,(1-x)^{2\alp -1} 
\over 32 \pi^2\,(\alp-1)(2\alpha-1)}
\ {x + {m \over M} \over \left(\lamR^2(x)\right)^{2\alpha - 2}}.
\ee
We note that these latter functions do not vanish at $x=0$, implying 
non-vanishing sum rules for $\gtwo$ and $\htwo$, in accordance
with Eqs. (\ref{bcsr}) and (\ref{bsr}), except if the quarks are massless. 

The functions $\gtwo$ and $\htwo$ are given by
\be
\gtwox = {\htwox \over 2} =
{N^2 \ar \ (1-x)^{2\alp -2} \over 32 \pi^2\,(\alp-1) (2\alpha-1)} \
{ 2( \alp -1) \left( x + {m \over M} \right) 
\left[ M^2 (1-x)^2 - M_R^2 \right] 
+ \left( 1 -2x -{m \over M} \right) \lamR^2(x) \over
\left( \lamR^2(x) \right)^{2 \alp -1}}.
\ee
We can directly check that Eqs.~(\ref{piet1}) and (\ref{piet2}) are satisfied.

Up to now, we have not specified flavor in the distributions. For the 
nucleon we only distinguished two types of distributions, $f_1^s$ and 
$f_1^a$, etc. Since spin 0 diquarks are in a flavor singlet
state and spin 1 diquarks are in a flavor triplet state, in order to combine to 
a symmetric spin-flavor wave function as demanded by the Pauli principle, 
the proton wave function has the well-known $SU(4)$ structure,
\be
\label{spinflavour}
|p \uparrow \rangle = {1 \over \sqrt{2}}\ \vert u \uparrow S_0^0\rangle +
{1 \over \sqrt{18}} \vert u \uparrow A_0^0\rangle
- {1\over 3} \ \vert u \downarrow A_0^1\rangle
- {1\over 3} \ \vert d \uparrow A_1^0\rangle
+ \sqrt{{2\over 9}} \ \vert d \downarrow A_1^1\rangle,
\ee
where $S$ ($A$) represents a scalar (axial vector) diquark and
the upper (lower) indices represent the projections of the spin (isospin)
along a definite direction. 
Since the coupling of the spin has already been included in the vertices, 
we need the flavor coupling
\be
\label{flavour}
|p \rangle = {1 \over \sqrt{2}}\ \vert u \ S_0\rangle +
{1 \over \sqrt{6}} \vert u \ A_0\rangle
- {1\over \sqrt{3}} \ \vert d \ A_1\rangle,
\ee
to find that for the nucleon the flavor distributions are
\bea
&&f_1^u = \frac{3}{2}\,f_1^s + \frac{1}{2}\,f_1^a, \\
&&f_1^d = f_1^a,
\eea
and similarly for the other functions.
The proportionality of the numbers is obtained from Eq.~(\ref{flavour}), 
while the overall factor is chosen to reproduce the sum rules for the number 
of up and down quarks if $f_1^s$ and $f_1^a$ are normalized to unity upon 
integration over $\bm p_T$ and $x$. This will fix the normalization $N$ 
in the form factor in Eq.~(\ref{ff}). 
Notice that the factors $a_s = 1$ and $a_a = -1/3$ in the distribution 
functions will produce different $u$ and $d$ weighting for unpolarized 
and polarized distributions. 
Further differences between $u$ and $d$ distributions can also be induced 
by different choices of $M_R$, $\Lambda$ or $\alpha$. 
We take for the nucleon $\alpha=2$ to reproduce the right large $x$ behavior 
of $f_1^u$, i.e.~$(1-x)^3$, as predicted by the Drell-Yan-West relation and 
reasonably well confirmed by data. 
We refrain from tuning the large $x$ behavior of $f_1^d$ to match the 
$(1-x)^4$ form indicated by data. Since $f_1^d$ is only
affected by vector diquarks, this could be easily obtained by choosing a
different form factor for the latter. We feel that this kind of fine-tuning 
would take things too far with the simple model we use. 
Similarly, we will only consider one common value for $\Lambda$.
We will, however, consider different masses for scalar and vector diquark
spectators. The color magnetic hyperfine interaction, held responsible 
for the nucleon-delta mass difference of 300 MeV, will also produce a 
mass difference between singlet and triplet diquark states. Neglecting 
dynamical effects, group-theoretical factors lead to a difference 
$M_a - M_s$ = 200 MeV \cite{close88}. 

Another important constraint comes from the axial charge of the nucleon,
given by
\be \label{axialcharge}
\gA = \int_0^1 dx \left[ g_1^u(x) - g_1^d(x) \right]
= \int_0^1 dx \left[ {3 \over 2} \ g_1^s(x)
- {1 \over 2} \ g_1^a(x) \right].
\ee

The sensitivity to the parameters $M_R$ and $\Lambda$ is best illustrated 
by considering some characteristic values.
We take a quark mass of 0.36 GeV (about one third of the 
average nucleon-delta mass), two different values
for $M_R$ (0.6 and 0.8 GeV) and three values for $\Lam$ (0.4, 0.5 and 0.6 GeV).
The distributions turn out to be insensitive to the value chosen
for the quark mass. In Table~\ref{sacomp} the values of some moments are given.

%
%
%
\begin{table}[b]
\caption{ \label{sacomp}
The second moment of $f_1$, $\langle x \rangle$ =
$\int dx\,xf_1(x)$ and the first moments $g_1$ = $\int dx\,g_1(x)$
and $h_1$ = $\int dx\, h_1(x)$ are given for two diquark masses
and for three values of $\Lambda$.}
\begin{tabular}{c|ccc|ccc}
& \multicolumn{3}{c}{$M_R = 0.6$ GeV} & \multicolumn{3}{c}{$M_R = 0.8$ GeV} \\
$\Lambda$ (GeV) & $\langle x\rangle^R$ & $g_1^R$ & $h_1^R$
& $\langle x\rangle^R$ & $g_1^R$ & $h_1^R$ \tstrut\\
\hline
0.4 & 0.366 & 0.923$\,a_R$ & 0.962$\,a_R$
    & 0.230 & 0.650$\,a_R$ & 0.825$\,a_R$ \tstrut\\
0.5 & 0.375 & 0.794$\,a_R$ & 0.897$\,a_R$
    & 0.256 & 0.527$\,a_R$ & 0.764$\,a_R$ \tstrut\\
0.6 & 0.384 & 0.671$\,a_R$ & 0.835$\,a_R$
    & 0.277 & 0.416$\,a_R$ & 0.708$\,a_R$ \tstrut 
\end{tabular}
\end{table}


%
%
%
\begin{table}[t]
\caption{ \label{dis-ud}
The second moment of $f_1$, $\langle x \rangle$ =
$\int dx\,xf_1(x)$ and the first moments $g_1$ = $\int dx\,g_1(x)$
and $h_1$ = $\int dx\, h_1(x)$ are given for $u$ and $d$ quarks in a proton
for three values of $\Lambda$.}
\begin{tabular}{c|ccc|ccc}
& \multicolumn{3}{c}{$u$-quark} & \multicolumn{3}{c}{$d$-quark} \\
$\Lambda$ (GeV) & $\langle x\rangle^u$ & $g_1^u$ & $h_1^u$
& $\langle x\rangle^d$ & $g_1^d$ & $h_1^d$ \tstrut\\
\hline
0.4 & $ 0.664 $ & $  1.277 $ & $  1.305 $
    & $ 0.230 $ & $ -0.217 $ & $ -0.275 $ \tstrut\\
0.5 & $ 0.690 $ & $  1.103 $ & $  1.218 $
    & $ 0.256 $ & $ -0.176 $ & $ -0.255 $ \tstrut\\
0.6 & $ 0.715 $ & $  0.937 $ & $  1.135 $
    & $ 0.277 $ & $ -0.139 $ & $ -0.236 $ \tstrut 
\end{tabular}
\end{table}

\begin{figure}[ht]
\begin{center}
\epsfig{file=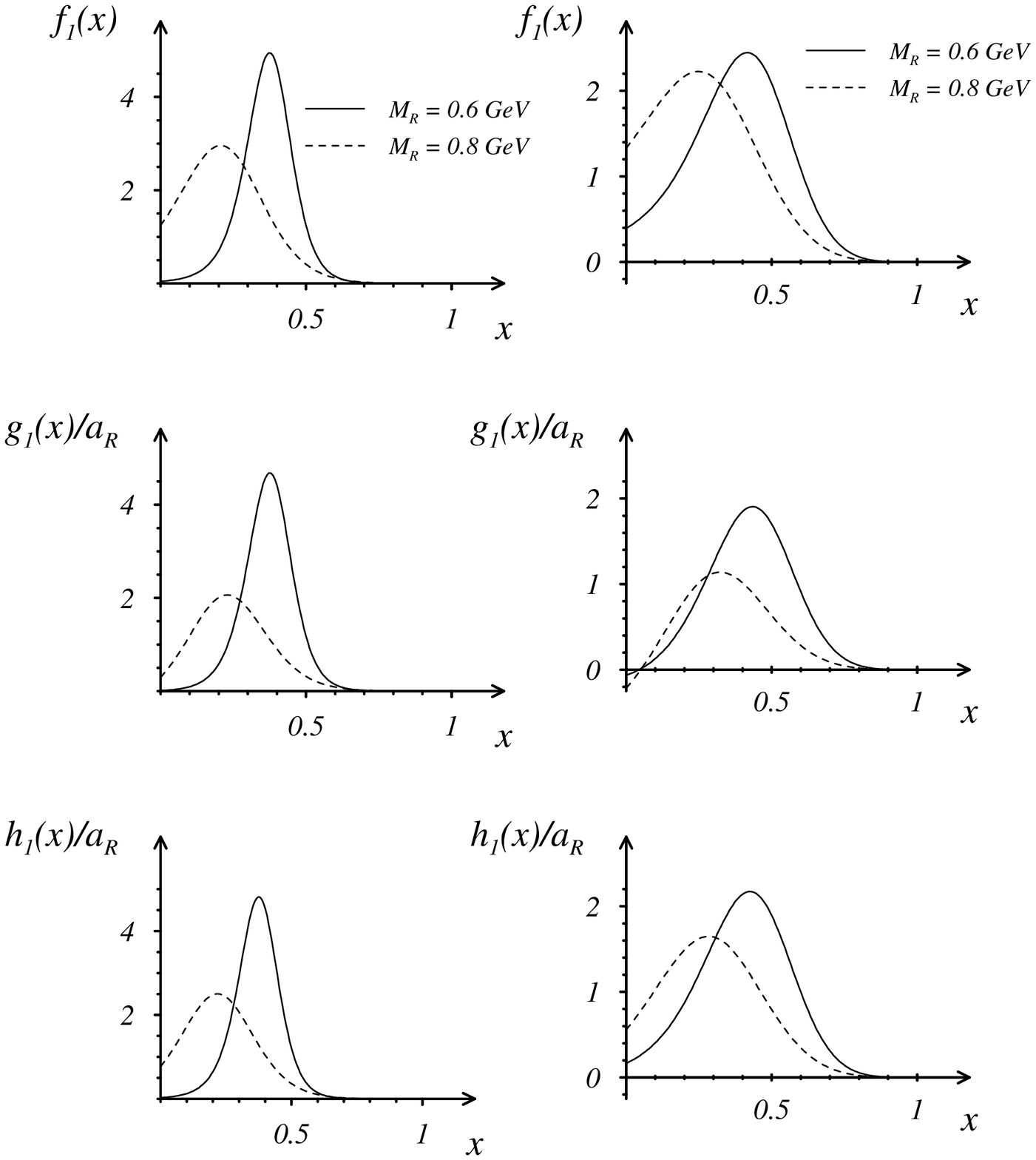,width=0.9\textwidth,angle=0}
\caption[dummy]{Twist two distributions for the nucleon.
The plots on the top represent $\fox$, the ones on the middle show $\gox / a_R$ 
and at the bottom we have $\hox /a_R$. 
The plots on the left correspond to $\Lam = 0.4$ GeV 
and the ones on the right to $\Lam = 0.6$ GeV.
The full line corresponds to $M_R = 0.6$ GeV and the dashed line to 
$M_R = 0.8$ GeV.}
\label{fgh}
\end{center}
\end{figure}

Fig.~\ref{fgh} shows the twist two distributions for different
values of the mass of the spectator and of the parameter $\Lam$. 
Clearly, $M_R$ dictates the position of the maximum, while $\Lam$ governs
the width of the distribution.
We can see that an increase of $M_R$ induces a shift on the peak of the 
valence distribution
$\fox$ towards lower values of $x$ and a decrease in its second moment.
In order to model sea quark distributions, one could introduce 
heavier spectators with four quarks or three quarks and one antiquark, 
and in this way satisfy the momentum sum rule at the model-scale.
In the absence of transverse momentum for the quarks he have 
$\fox = \gox /a_R = \hox/a_R$.

%
%

In Table \ref{dis-ud} we have given the values of some moments for $u$ and
$d$ quarks in the proton using $M_s = 0.6$ GeV and $M_a = 0.8$ GeV.We now use
the axial charge of the nucleon, $g_A=g_1^u-g_1^d$, to find the most suitable 
values for $\Lam$. The value $\Lam = 0.5$ GeV gives $\gA = 1.28$, close to the
experimental value.

%
%
\begin{figure}[h]
\begin{center}
\epsfig{file=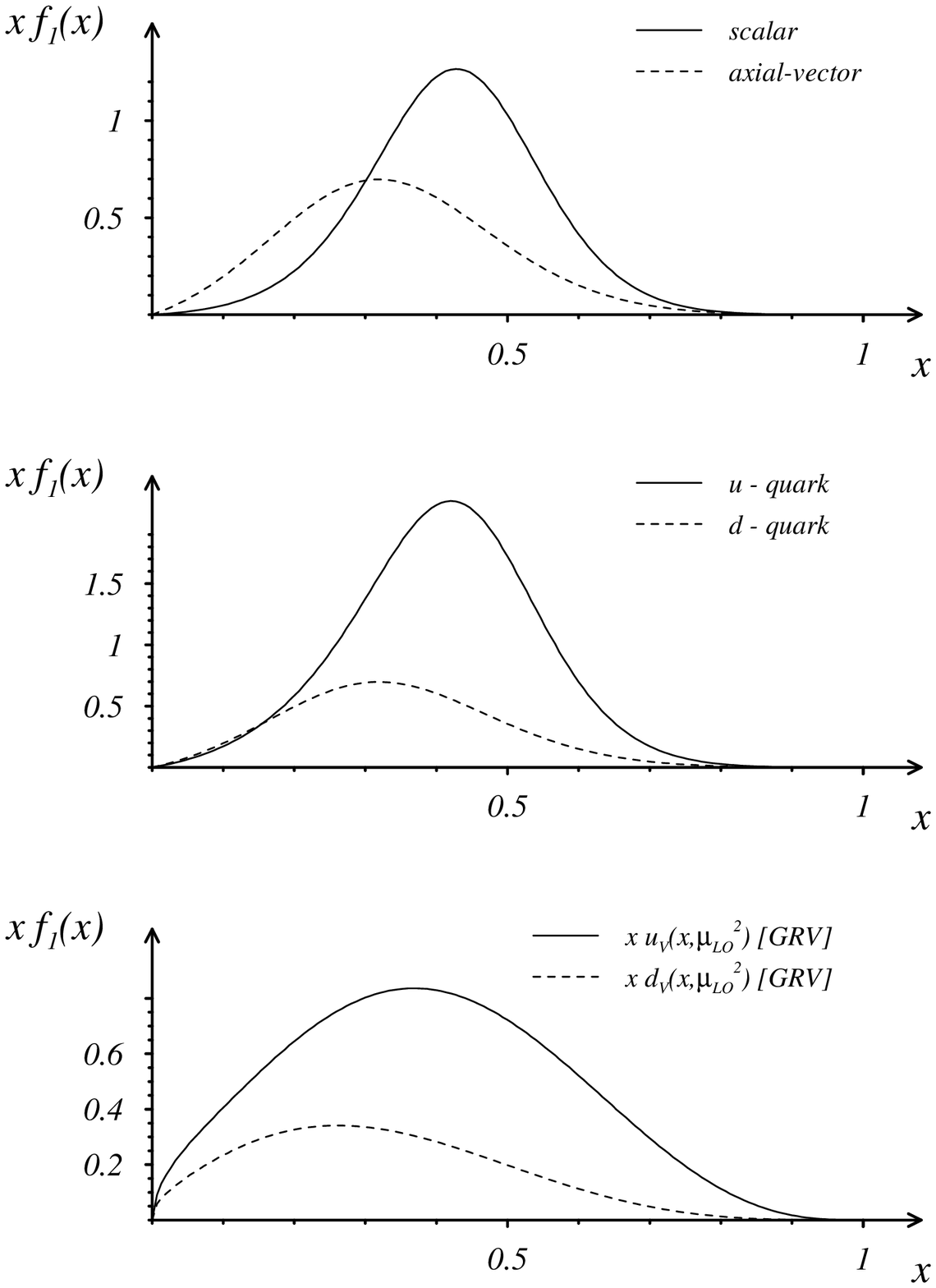,width=0.9\textwidth,angle=0}
\caption[dummy]{Twist two distributions for the nucleon.
The plot at the top shows $x f_1^s(x)$ (full line) and $x f_1^a(x)$ 
(dashed line) for $M_s=0.6$ GeV, $M_a = 0.8$ GeV and $\Lam=0.5$ GeV.
The plot on the middle shows $x f_1^u(x)$ (full line) and $x f_1^d(x)$ 
(dashed line) for the same values of the parameters.
The third plot shows the low scale ($\mu^2 = 0.23$ GeV$^2$)
valence distributions of Gl\"{u}ck, Reya and Vogt \cite{grv95}.}
\label{xf}
\end{center}
\end{figure}

\begin{figure}[ht]
\begin{center}
\epsfig{file=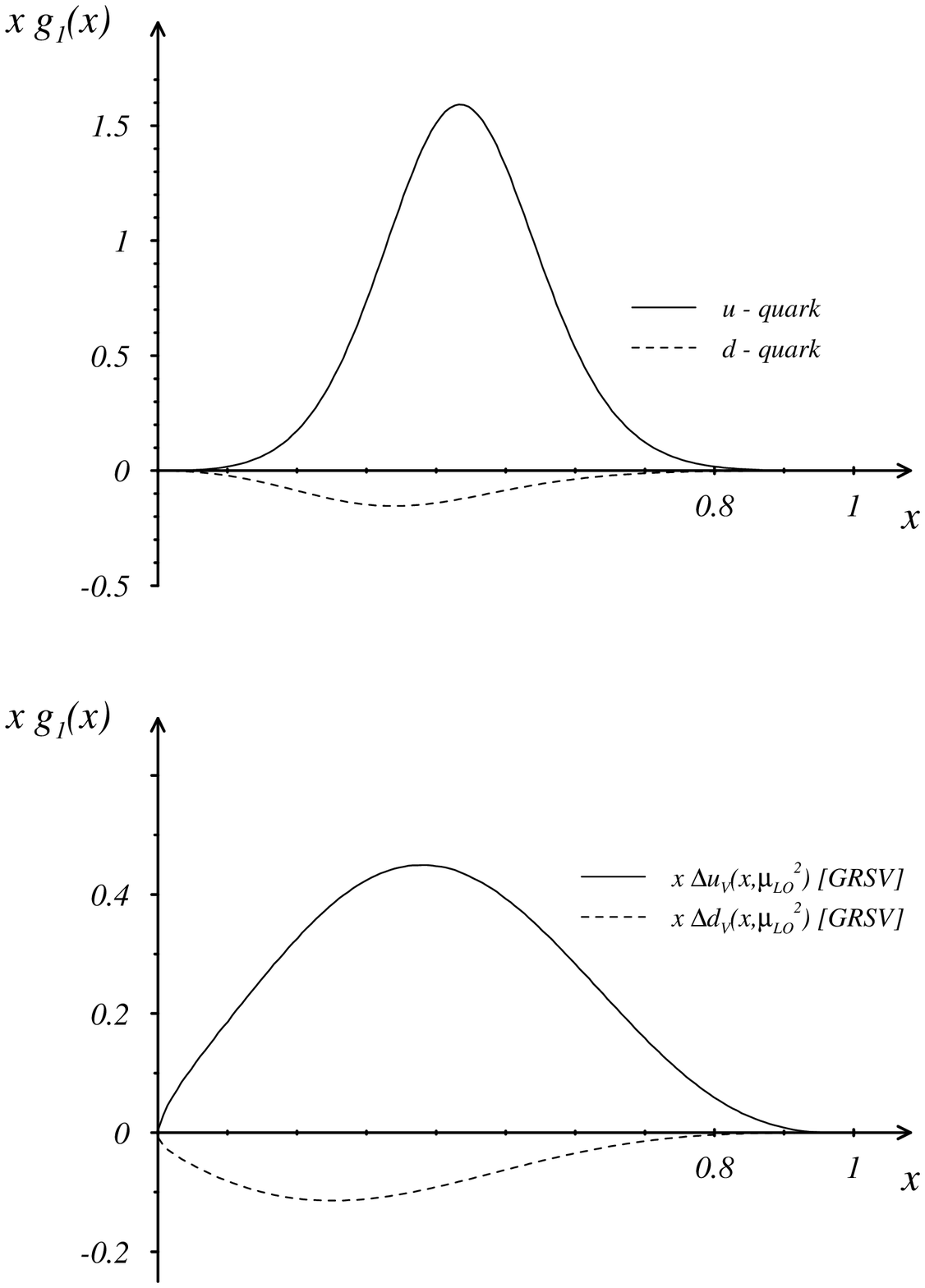,width=0.9\textwidth,angle=0}
\caption[dummy]{Polarized proton distributions $g_1^u(x)$ and $g_1^d(x)$.
The first plot shows our estimates for $x g_1^u(x)$ (full line)
and $x g_1^d(x)$ (dashed line) for $\Lam = 0.5$ GeV, $M_s = 0.6$ GeV 
and $M_a = 0.8$ GeV. The second plot shows the low scale 
$\mu_{LO}^2 = 0.23$ GeV$^2$ parametrization of
Gl\"{u}ck, {\it et al.} \cite{grsv96} for the same functions}.
\label{xg}
\end{center}
\end{figure}

\begin{figure}[ht]
\begin{center}
\epsfig{file=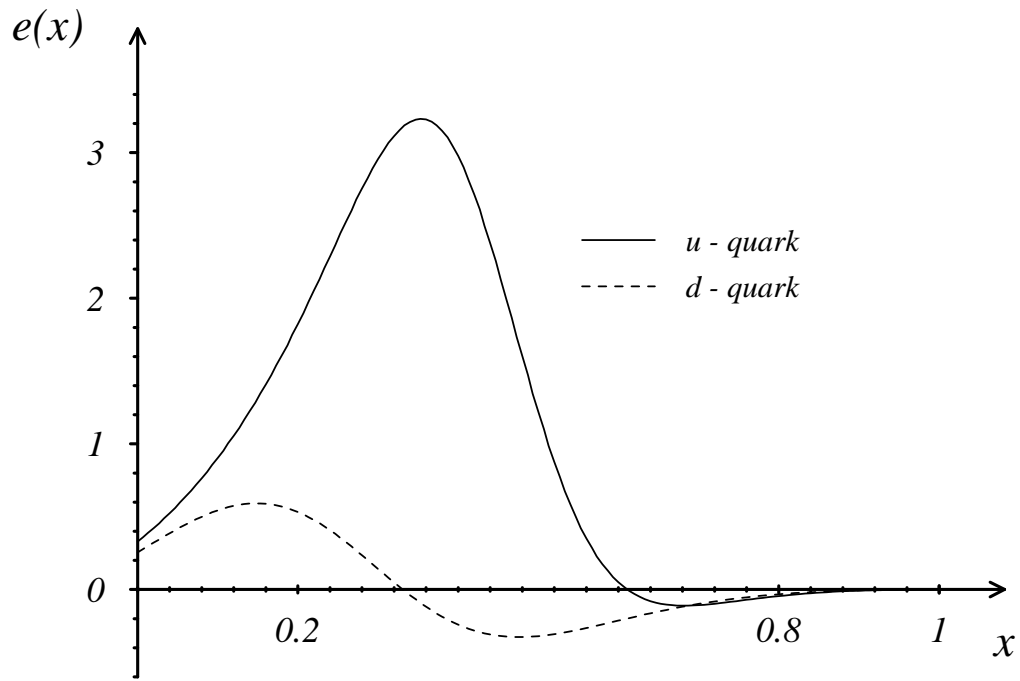,width=0.9\textwidth,angle=0}
\caption[dummy]{Distributions $e^u(x)$ (solid line) and $e^d(x)$ (dotted line) 
for $M_s=0.6$ GeV, $M_a = 0.8$ GeV and $\Lam=0.5$ GeV.}
\label{e}
\end{center}
\end{figure}

\begin{figure}[ht]
\begin{center}
\epsfig{file=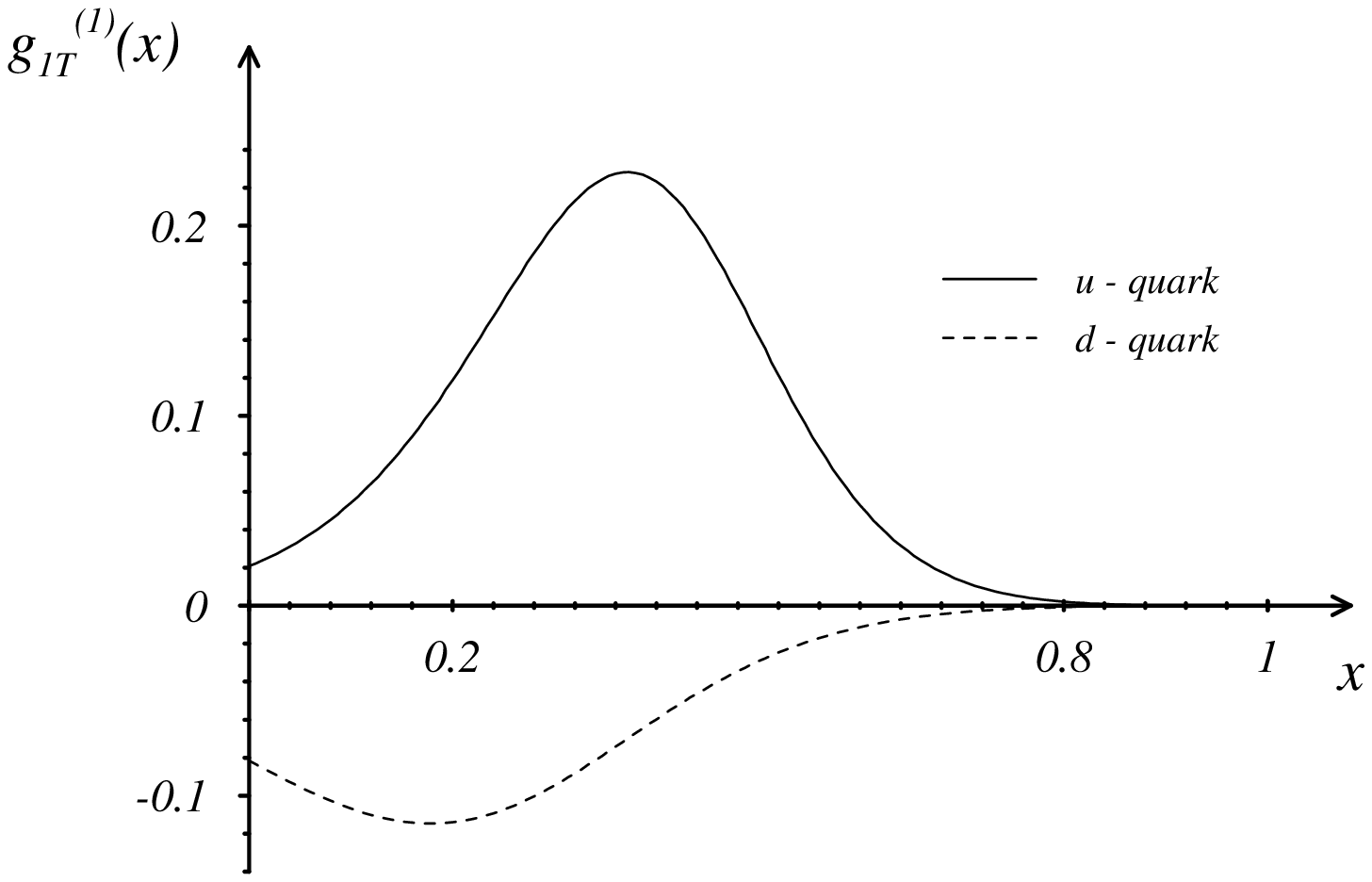,width=0.9\textwidth,angle=0}
\caption[dummy]{Distributions $g_{1T}^{(1)u}(x)$ (solid line) and 
$g_{1T}^{(1)d}(x)$ (dashed line) 
for $M_s=0.6$ GeV, $M_a = 0.8$ GeV and $\Lam=0.5$ GeV.}
\label{g1T1}
\end{center}
\end{figure}

\begin{figure}[h]
\begin{center}
\epsfig{file=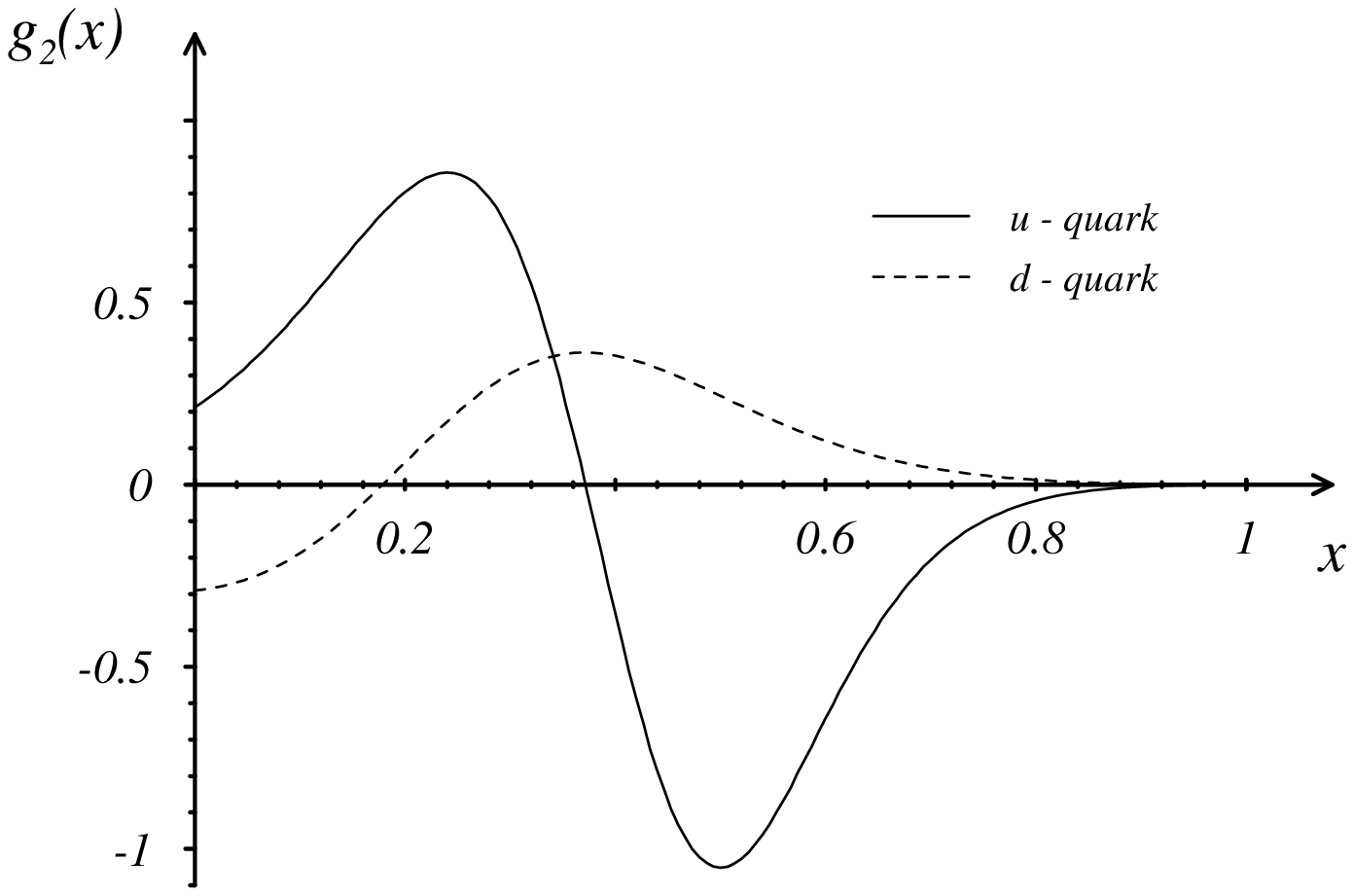,width=0.9\textwidth,angle=0}
\caption[dummy]{Distributions $\gtwo^u(x)$ (solid line) and 
$\gtwo^d(x)$ (dashed line) 
for $M_s=0.6$ GeV, $M_a = 0.8$ GeV and $\Lam=0.5$ GeV.}
\label{g2}
\end{center}
\end{figure}

Fig. \ref{xf} shows the distribution $\fox$ multiplied by $x$.
We find a satisfactory qualitative agreement with the valence 
distributions of Gl\"{u}ck, Reya and Vogt (GRV)
calculated at the low scale $\mu^2_{LO} = 0.23$ GeV$^2$ \cite{grv95}.
For $u$ and $d$ quarks, the first moment of $\fox$ is clearly 
larger in our model, which would imply that our results describe the 
nucleon at an even lower scale than GRV.
Fig.~\ref{xg} shows the distributions $g_1^u(x)$ and $g_1^d(x)$ 
multiplied by $x$ for the values 
$M_s=0.6$ GeV, $M_a = 0.8$ GeV and $\Lam=0.5$ GeV.
Again, we find a qualitative agreement with the polarized valence 
distributions of Gl\"{u}ck {\it et al.} \cite{grsv96}.
Using the same parameters, we can obtain higher twist distributions. 
Fig. \ref{e} shows the twist three distributions $e^u(x)$ and $e^d(x)$,
Fig. \ref{g1T1} shows the distributions $g_{1T}^{(1)u}(x)$ 
and $g_{1T}^{(1)d}(x)$, while 
in Fig. \ref{g2} we have the distributions $\gtwo^u(x)$ and $\gtwo^d(x)$.
We find a small violation of the Burkhardt-Cottingham sum rule, in 
agreement with Eq.~(\ref{bcsr}), due to the non-zero value of $g_{1T}^{(1)}(0)$.

%
%

%
%

%
%

%
%

At this point it is important to realize that in this model 
there are no antiquarks.
This means that the distribution functions are zero for $x<0$, due to the
symmetry properties of the matrix elements involved in the calculation.
Therefore, $C$-even and $C$-odd sum rules are equal.


\subsection{Distribution functions for the pion}

\begin{figure}[h]
\begin{center}
\epsfig{file=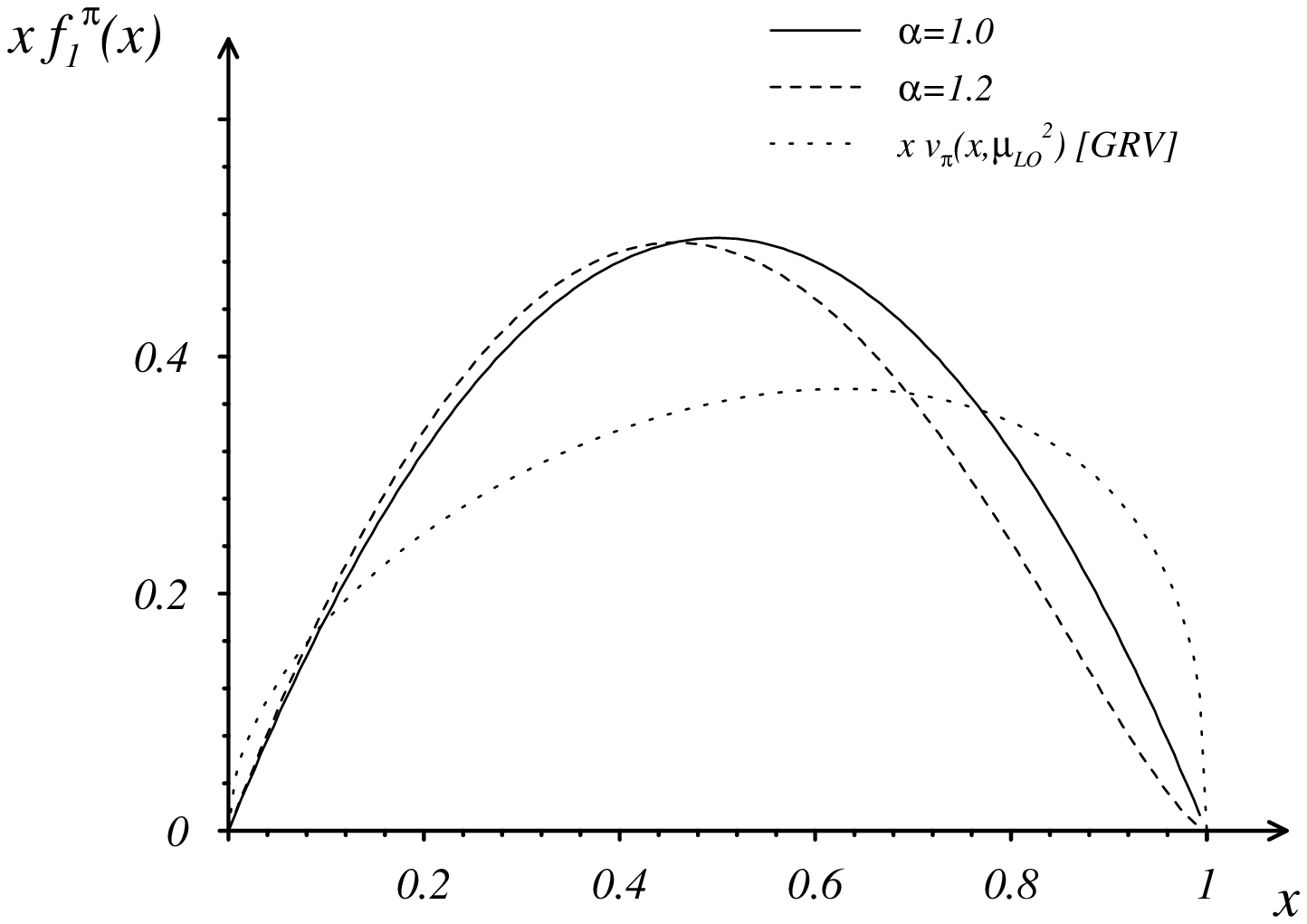,width=0.9\textwidth,angle=0}
\caption[dummy]{Momentum distribution $x \fox$ for the pion
for $\alp = 1.0$ (solid line) and $\alp = 1.2$, $\Lam= 0.4$ GeV (dashed line).
The dotted line represents the GRV low scale parametrization \cite{grv92}.}
\label{xfpi}
\end{center}
\end{figure}

The expressions for the 
distribution functions for the pion are the same as those for
the nucleon with $a_R=0$ and making the replacement
$M_R \rarrow m$. Only spin independent functions will remain.
For $\alp = 1$ we have $\fox = 2 (1-x)$, showing the power law behavior 
expected from simple counting rules. 
The apparent singularity caused by the factor $(\alp-1)$ that enters in
the denominator can be avoided including this factor in the normalization $N$.
In this case we find the symmetry $x \leftrightarrow (1-x)$ for $x \fox$.
For values of $\alp$ different from 1 
the functions depend on the parameter $\Lam$, which
is constrained by Eq.~(\ref{condition1}). For
$\Lam = 0.4$ GeV, the distribution $\fox$ is shown in Fig.~\ref{xfpi}
for two values of $\alp$ and compared with the parametrization of the
leading order valence distributions of Gl\"{u}ck, Reya and Vogt at the low scale
$\mu^2_{LO} = 0.25$ GeV$^2$ \cite{grv92}.

In this case the vertex gives immediately identical antiquark distribution
or, equivalently, a contribution for negatives values of $x$.

%
%
%


\subsection{Fragmentation functions for the nucleon}

The assumed form of the quark-diquark-nucleon vertex also allows the
calculation of the fragmentation functions for the nucleons. We can 
use the reciprocity relation mentioned at the end of section III to
obtain the analytic expressions 
for $D_1(z, -z \ktns)$, etc., and, after integration
over $\kpns = -z \ktns$, the expressions for $D_1(z)$ etc. 
In this case we introduce the function $\lamR$ through the relation

\be
k^2 - \Lam^2 = {\kts + \lamR^2 (1/z) \over 1 -z}.
\ee

For example, the unpolarized fragmentation function reads

\be
D_1(z,z^2 \kts) = {N_F^2 (1-z)^{2\alp-1} \over 16 \pi^3 z^{2\alp}} \ 
{\left( {M \over z} + m  \right)^2 + \kts \over 
\left( \kts + \lamR^2 \left( {1 \over z} \right) \right)^{2 \alp}},
\ee

\nin and, after integration over the transverse momentum,

\be
z \Doz = {N_F^2 \, z^{2\alp -1} \, (1- z)^{2 \alp -1} \over 
32 \pi^2\,(\alpha-1)(2\alpha-1)} 
\ {2(\alpha-1)\, (M + mz)^2 
+ z^2 \, \lamR^2 \left( {1 \over z} \right) \over 
\left(z^2 \lamR^2 \left( {1 \over z} \right) \right)^{2\alpha - 1}}.
\ee

The factor $N_F$ is a normalization constant.
Distinguishing $D_1^s (D_1^a)$ as the fragmentation functions for a quark
into a nucleon and an anti-S diquark (anti-$A$ diquark), one finds
\bea \label{dup}
D_1^{u \rarrow p} & = & {3 \over 2} \ D_1^s + \half D_1^a, \\
D_1^{u \rarrow n} & = & D_1^{d \rarrow p} = D_1^a, \label{dun}
\eea

\nin and similarly for $G_1$ and $H_1$.
In this case there is no sum rule to fix the normalizations of $D_1^s$ 
and $D_1^a$. In the $SU(4)$ symmetric limit $(D_1^s = D_1^a)$, one finds
the expected ratio $D_1^{u \rarrow p}/D_1^{d \rarrow p} = 2$. We will
introduce the scale invariant quantities

\be \label{epsilonr}
\epsR = \int_0^1 dz \,z \,D_1^R(z),
\ee

\nin and express our results with the help of these quantities.
In Table~\ref{sacomp2} the values of some moments are given for
a quark mass of 0.36 GeV, two different values
of $M_R$ (0.6 and 0.8 GeV) and three values of $\Lam$ (0.4, 0.5 and 0.6 GeV.)
By normalizing $\eps^s = \eps^a = 1$, we obtain the results for $u$ and $d$
quarks given in Table~\ref{sacomp3}.

%
%
%
\begin{table}[t] 
\caption[dummy2]{\label{sacomp2}
The first two moments of the unpolarized fragmentation function 
$D_1^R$ ($N^R=\int D_1^R(z)\,dz$ and $\epsilon^R$) 
and the ratios $G_1^R/N^R=\int G_1^R(z)\,dz/N^R$ and 
$H_1^R/N^R=\int H_1^R(z)\,dz/N^R$ for different values of $\Lambda$
and $M_R$.}
\begin{tabular}{c|cccc|cccc}
& \multicolumn{4}{c}{$M_R = 0.6$ GeV} & \multicolumn{4}{c}{$M_R = 0.8$ GeV} \\
$\Lambda$ (GeV) & $N^R$ & $\epsilon^R$ & $G_1^R/N^R$ & $H_1^R/N^R$
& $N_R$ & $\epsilon^R$ & $G_1^R/N^R$ & $H_1^R/N^R$ \tstrut\\
\hline
0.4 	& $1.908\;\epsR$		& $2.732 \times 10^{-5}$ 	
	& $0.628\;\ar $          	& $0.814\;\ar $
	& $2.112\;\epsR$	 	& $3.785 \times 10^{-4}$ 	
	& $0.538\;\ar $          	& $0.769\;\ar $ \tstrut\\
0.5 	& $1.901\;\epsR$     		& $4.808 \times 10^{-4}$
	& $0.639\;\ar $          	& $0.820\;\ar $
	& $2.107\;\epsR$   		& $1.345 \times 10^{-3}$
	& $0.548\;\ar $          	& $0.774\;\ar $ \tstrut\\
0.6 	& $1.891\:\epsR$	   	& $2.429 \times 10^{-3}$
	& $0.654\;\ar $          	& $0.827\;\ar $
	& $2.099\;\epsR $	    	& $3.732\times 10^{-3}$
	& $0.561\;\ar $          	& $0.781\;\ar $ \tstrut
\end{tabular}
\end{table}
%

%
%

\begin{table}[b]
\caption[dummy3]{\label{sacomp3}
The first two moments of the $u$ and $d$ quark combinations of 
Eqs.~(\ref{dup}) and (\ref{dun})
and the ratios $G_1^q/N^q=\int G_1^q(z)\,dz/N^q$ and 
$H_1^q/N^q=\int H_1^q(z)\,dz/N^q$ for fragmentation into protons.}
\begin{tabular}{c|cccc|cccc}
& \multicolumn{4}{c}{$u$-quark} & \multicolumn{4}{c}{$d$-quark} \\
$\Lambda$ (GeV) & $N^u$ & $\epsilon^u$ & $G_1^u/N^u$ & $H_1^u/N^u$
& $N^d$ & $\epsilon^d$ & $G_1^d/N^d$ & $H_1^d/N^d$ \tstrut\\
\hline
0.4 	& $3.917$	& $2$ 	
	& $0.410$      	& $0.525$
	& $2.112$ 	& $1$ 	
	& $-0.179$     	& $-0.256$ \tstrut\\
0.5 	& $3.904$     	& $2$
	& $0.418$      	& $0.529$
	& $2.107$	& $1$
	& $-0.183$     	& $-0.258$ \tstrut\\
0.6 	& $3.886$	& $2$
	& $0.427$      	& $0.533$
	& $2.099$    	& $1$
	& $-0.187$	& $-0.260$ \tstrut
\end{tabular}
\end{table}

\begin{figure}[h]
\begin{center}
\epsfig{file=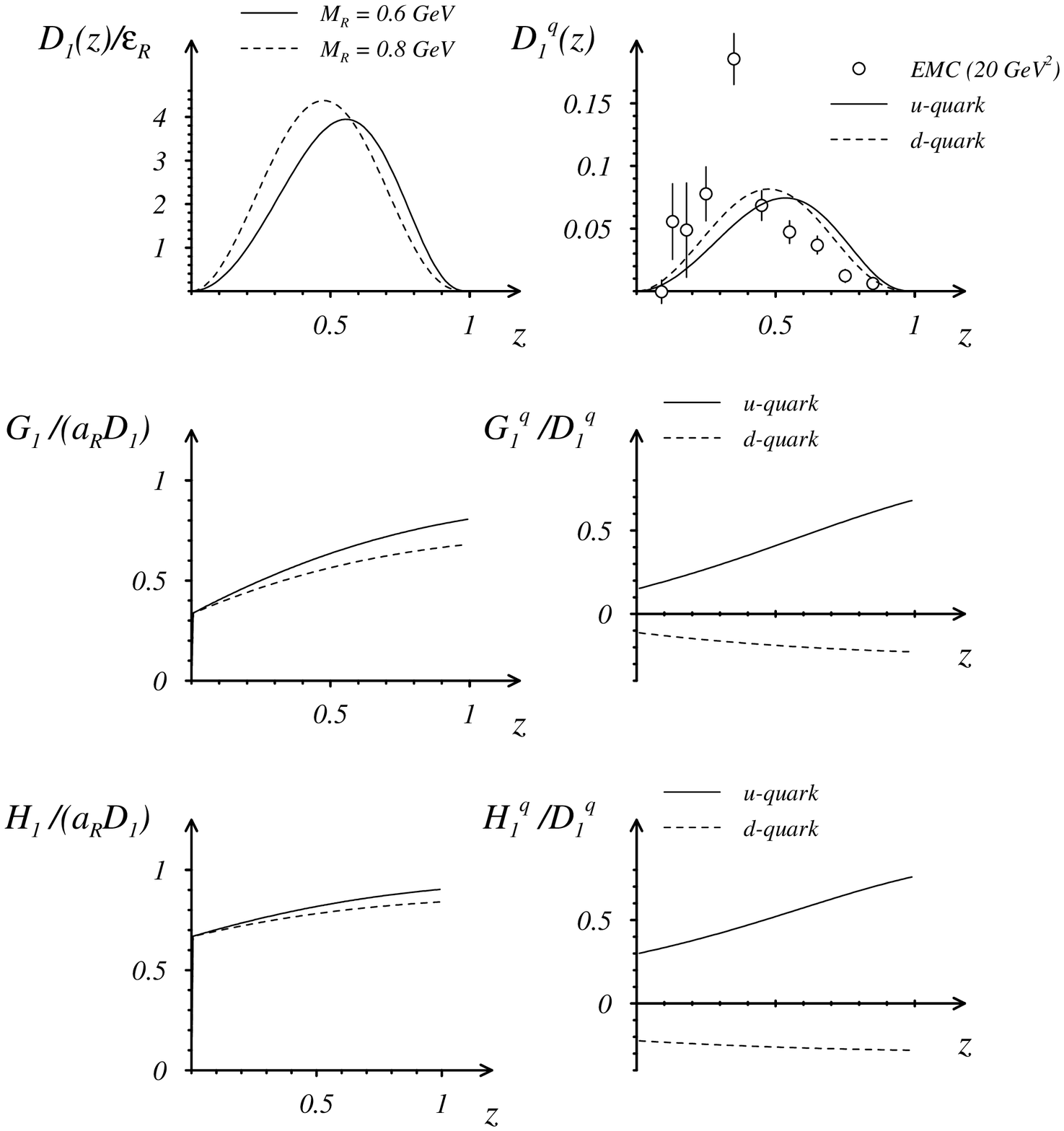,width=0.9\textwidth,angle=0}
\caption[dummy4]{
Twist two fragmentation functions. The plots on the left show
$D_1^R(z)/\epsilon^R$ (with $\epsilon^R$ defined according to 
Eq.~(\ref{epsilonr})) and
the ratios $G_1(z)/(a_RD_1^R(z))$ and $H_1(z)/(a_RD_1^R(z))$ for $M_R = 0.6$
GeV and $M_R = 0.8$ GeV ($\Lambda = 0.5$ GeV). The plots on the right show the
$u$ and $d$ quark results obtained with $\epsilon^s=\epsilon^a=1$. For
comparison with data our results
for $D_1^q(z)$ are rescaled (see text). Data for the difference
$(D_{1}^{u\to p}(z)-D_{1}^{u\to\bar p}(z))$ are taken from \cite{EMC}.}
\label{DGH}
\end{center}
\end{figure}

In Fig.~\ref{DGH} we show our results for the unpolarized fragmentation function
$D_1$ and the ratios of polarized to unpolarized functions $G_1(z)/D_1(z)$ 
and $H_1(z)/D_1(z)$. Since the dependence on the parameter $\Lambda$ turns out
to be weak, we display results for the choice $\Lambda=0.5$ GeV only. On the
left hand side of the figure the two spectator masses ($M_R=0.6$ GeV and
$M_R=0.8$ GeV) are compared. On the right hand side we show the results for 
the $u$ and the $d$ quark fragmentation functions as defined by 
Eqs.~(\ref{dup}) and (\ref{dun}). To allow a comparison with data \cite{EMC}
we fixed the normalization such that the second moment 
$\epsilon^{q\to p}=\int dz\,z\,D_1^q(z)$ takes the value
\begin{equation}
\int dz\,z\,\left(D_1^{u\to p}(z)-D_1^{u\to\bar p}(z)\right)\approx 0.019, 
\end{equation}
which is our (rough) estimate for the second moment obtained from the EMC 
data. We compare our result for $D_1^u(z)$ to the difference
\begin{equation}
\label{valcomb}
D_1^{u\to p}(z)-D_1^{u\to\bar p}(z)
=D_1^{u\to p}(z)-D_1^{\bar u\to p}(z),
\end{equation}
since it is the appropriate combination for comparison with a model
involving only valence quarks, although in our 
model $D_1^{u\to\bar p}$ is zero by construction (as are all
so-called unfavored fragmentation functions). 
Furthermore, since in the EMC analysis the assumption 
$D_1^{u\to p}(z)=D_1^{d\to p}(z)$ was made, we compare our result for 
$D_1^d(z)$ to just the same combination of Eq.~(\ref{valcomb}). The ratios of 
polarized to unpolarized fragmentation functions, $\Goz/\Doz$ and $\Hoz/\Doz$,
are given for the $q=u,d$ as well; normalization factors drop from the ratios.

%
%


\subsection{Fragmentation functions for the pion}

\begin{figure}[h]
\begin{center} 
\epsfig{file=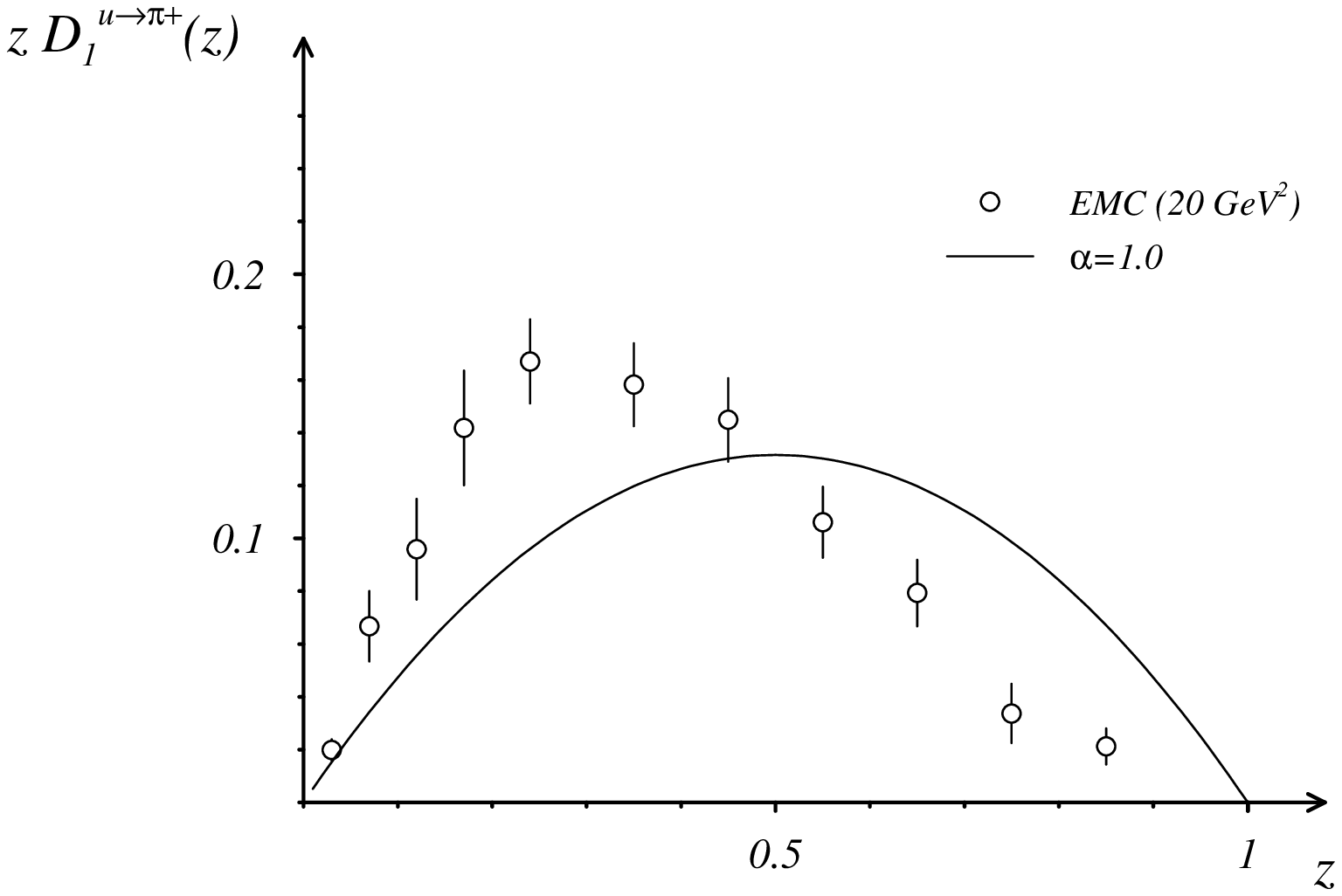,width=0.9\textwidth,angle=0}
\caption[dummy5]{
The pion fragmentation function $z\,D_1^{u\to{\pi^+}}(z)$ (rescaled, 
see text) for $\alp = 1.0$ is compared to the
EMC data from Ref.~\cite{EMC}. Data are shown for the difference
($z\,D_1^{u\to{\pi^+}}-z\,D_1^{u\to{\pi^-}}$).}
\label{zDpi}
\end{center}
\end{figure}

Our results for the fragmentation function of a quark to a pion are shown 
in Fig.~\ref{zDpi}. We display $zD_1^{u\to\pi^+}(z)$ which is rescaled such 
that the second moment 
$\epsilon^{q\to \pi}=\int dz\,z\,D_1^{u\to\pi^+}(z)$ equals the value
of the valence combination 

\begin{equation}
\int dz\,z\,\left(D_1^{u\to{\pi^+}}(z)-D_1^{u\to{\pi^-}}(z)\right)
\approx 0.088, 
\end{equation}

\nin our estimate for the second moment obtained from the corresponding 
EMC data \cite{EMC} (more recent parametrizations available 
for the combination ($D_1^{u \to {\pi^+}} + D_1^{u \to {\pi^-}}$)
\cite{bkk95} agree with the EMC data).

Note that in our calculation all favored fragmentation functions are identical
\begin{equation}
D_1^{u\to{\pi^+}}(z) = D_1^{\bar d\to{\pi^+}}(z)=
D_1^{d\to{\pi^-}}(z) = D_1^{\bar u\to{\pi^-}}(z),
\end{equation}
while all unfavored fragmentation functions 
have not been considered in our approach:

\begin{equation}
D_1^{d\to{\pi^+}}(z)=D_1^{\bar u\to{\pi^+}}(z)=
D_1^{\bar d\to{\pi^-}}(z)=D_1^{u\to{\pi^-}}(z)=0.
\end{equation}
The latter property has to be contrasted with the experimental observation 
that unfavored fragmentation functions can be as large as the favored ones
for small $z$, are suppressed by a factor of about 2 for $0.4\le z\le 0.6$ 
and even stronger suppressed for large $z$ \cite{EMC}. This observation 
holds true for both nucleons and pions.

%
%


\subsection{Summary}

In this paper we combined the representation of distribution and fragmentation
functions in terms of non-local operator expectation values
with a simple spectator model. This amounts to saturating the antiquark-hadron
intermediate state with one single state of definite mass. With an
effective vertex that connects to this state, containing a form factor,
we can calculate all the independent amplitudes that are allowed for the 
non-local operator expectation values after imposing constraints of
Lorentz invariance and invariance under parity and time-reversal
operations. Exploiting the explicit expressions of distribution and
fragmentation functions in terms of  those amplitudes, several relations between
$p_T$-integrated distribution (or fragmentation) functions arise. 

For nucleons and pions we have obtained expressions for the distribution
and fragmentation functions within our approach. 
Flavor charges and axial vector charge 
served to fix the free parameters of the model in the case of the distribution
functions. For the fragmentation functions we utilized the same set of
parameters except for the overall normalization which in this case is not
constrained by a number sum-rule. Considering all Dirac projections in leading
and subleading order (in an expansion in 1/Q) we were able to give estimates
for the polarized and unpolarized cases, including
subleading (higher twist) functions. 
The latter lack a simple probabilistic interpretation, but are 
well defined as projections of non-local operators.

By comparing our expressions to available parametrizations at `low
(hadronic) scales' and to some experimental data we find 
that we can obtain reasonable qualitative agreement for the
unpolarized distribution function $f_1(x)$, the longitudinal spin-distribution
$g_1(x)$ and with the unpolarized fragmentation function
$D_1(z)$ for both nucleons and pions.
These findings give confidence that the estimates obtained
for the `terra incognita' functions (transverse spin distributions,
longitudinal and transverse spin fragmentations and subleading functions)
provide a reasonable 
estimate of the order of magnitude of the functions and their
(large) $x$ behavior despite the simplicity of the model. 
The comparison to the available parametrizations and
experimental data gives an indication of the level of accuracy our
estimates can reach, keeping in mind that we have excluded the
sea-quark and gluon sectors, and evolution.

\acknowledgments

This work is supported by the Foundation for Fundamental Research on Matter
(FOM), the National Organization for Scientific Research (NWO) and the Junta
Nacional de Investiga\c{c}\~{a}o Cient\'{\i}fica (JNICT, PRAXIS XXI).


\end{document}